\documentclass[12pt,preprint]{aastex}

\slugcomment{submitted to \apj}
\shorttitle{Bulges and disks in 3CR galaxies}
\shortauthors{Donzelli et al.}

\begin{document}

\title{Bulges  and disks  in the  Host  Galaxies of  low redshift  3CR
Sources:    a    near-IR    view    of   their    radial    brightness
profiles\altaffilmark{1}}


\author{
Carlos J. Donzelli\altaffilmark{2,3,4,5},
Marco Chiaberge\altaffilmark{3,6},
F. Duccio~Macchetto\altaffilmark{3,7},
Juan P. Madrid\altaffilmark{3},      
Alessandro Capetti \altaffilmark{8}, and
Danilo Marchesini \altaffilmark{9}
}

\altaffiltext{1}{Based on  observations made with  the NASA/ESA Hubble
Space Telescope (HST),  obtained at  the Space Telescope  Science Institute,
which is operated  by the Association of Universities  for Research in
Astronomy, Inc., under NASA  contract NAS 5-26555.}

\altaffiltext{2}{ Gemini Fellow }

\altaffiltext{3}{ Space  Telescope Science Institute,  3700 San Martin
Drive, Baltimore, MD 21218, USA}

\altaffiltext{4}{ IATE, Observatorio Astron\'omico, UNC, Laprida 854, C\'ordoba, Argentina }

\altaffiltext{5}{ CONICET, Rivadavia 1917, Buenos Aires, Argentina }

\altaffiltext{6}{On leave From INAF, Instituto di Radioastronomia, Via
P.  Gobetti 101, Bologna, Italy, 40126-I.}

\altaffiltext{7}{ Research \& Space Science Department,
European   Space   Agency,    ESTEC,   Noordwijk,   The   Netherlands }

\altaffiltext{8}{ INAF - Osservatorio Astronomico di Torino, Strada Osservatorio
20, 10025 Pino Torinese, Italy }

\altaffiltext{9}{ Department of Astronomy, Yale University, New Haven, CT 06520, USA }


\begin{abstract}

We  analyze  the  near-infrared  luminosity profiles  and  photometric
parameters of the host galaxies  of 3CR radio sources with $z<0.3$,
to investigate  their physical  nature.  Our  sample  includes 82
galaxies, of  which 22 (27\%) are  FR~Is  and 60 (73\%)  are  FR~IIs.
Using near-infrared data taken both with NICMOS onboard the Hubble Space Telescope
and  from the ground with the Telescopio Nazionale Galileo,
we find that  luminosity profiles are very well  described by a single
S\'ersic law in 52\% of the cases and for the remaining objects (48\%)
it is necessary to include an exponential profile, which could indicate the
presence of  a disk.  The average  bulge to disk  luminosity ratio for
the  galaxies  is  $(b/d)  \sim  1.1$.  
The  analysis   of  the  photometric
parameters  of the  sub samples  indicates  that FR~Is and FR~IIs  show
rather similar bulges in terms of effective surface magnitude, effective radius, 
and S\'ersic index.
On  the other  hand, disks  in FR~Is  and FR~IIs  hosts show,  on average,
different properties. Central surface  magnitudes are dimmer and scale
lengths are  greater by a  factor of 2 in FR~Is  when compared
to FR~IIs.\\
We also estimate the black
hole mass associated with each galaxy using two  different methods that
claim tight  correlations between the black hole  mass ($M_{BH}$) with
the  infrared bulge  luminosity  ($L_{bulge}$) and  with the  S\'ersic
index ($n$). Our data indicate that masses obtained through these two
methods  show   a  high  dispersion  and   $M_{BH}$  obtained  through
$L_{bulge}$ are systematically  higher (by a factor of  $\sim$ 3) than
those obtained  using $n$.  This result may  reflect  the
fact  that for our sample  galaxies we  do not  find  any correlation
between $L_{bulge}$ and $n$.

\end{abstract}



\section{Introduction}
Radio galaxies  are a peculiar  set of galaxies  which give rise  to a
number of questions regarding their  nature: What kind of effects turn
these  galaxies into  radio  galaxies? Are  these  effects related  to
galaxy  morphology  or  galaxy  environment? What  are  the  relations
between  the  properties  of  radio  galaxies and  the  normal  galaxy
population? Are  these properties  a result of  the radio  activity or
vice versa?  In  the framework of the AGN  unification scheme (Urry \&
Padovani 1995), radio galaxies are associated with quasars.  Therefore
the  study  of such  objects  plays  an  important role  to  constrain
physical models  of quasar evolution (Kauffmann \&  Haehnelt 2000), to
further test  the unified models, and  to explore the  origin of radio
loudness (Blandford 2000).\\ 

Radio galaxies are usually  classified into two morphological classes,
edge-darkened  FR~Is  and  edge-brightened  FR~II (Fanaroff  \&  Riley
1974), and this dichotomy  corresponds to a (continuous) transition in
radio  power,  which occurs  at  $L_{178}  \sim  2 \times 10^{26}$  W
Hz$^{-1}$.   In   modern  high   resolution  radio  maps,   FR~Is  are
charachterized by bright jets near the radio ``core'' (centered on the
host galaxy  nucleus), while FR~IIs have faint  jets (mostly one-sided
as a result  of relativistic beaming effects) coupled  with bright hot
spots in the lobes, which mark  the location where the jet impacts the
ISM.

Optical studies  of low-redshift radio galaxies have  shown that radio
galaxies of  both classes are almost invariably  associated with giant
ellipticals, are generally found in regions of enhanced galaxy density
such as clusters or dense  groups, and often show signs of interaction
(e.g.   Zirbel  1997, Gonzalez-Serrano  et  al  1993).  However,  some
differences between  FR~Is and FR~II  seem to be present.   The most
striking difference is related to  the emission line properties of the
two classes:  while most  (if not all)  FR~Is only show  weak (narrow)
emission lines, FR~II can be distinguished into different sub-classes,
according  to  their  spectral  properties  (e.g.   Laing  1994).   In
particular, the role  of a subclass of low  excitation FR~II (LEG) has
still to be fully established:  the central engine and the environment
of  LEG appears  very  similar to  that  of FR~Is  (e.g. Chiaberge  et
al. 2002,  Hardcastle \&  Worrall 2000), while  the radio
power and morphology are typical of FR~II.

While the  above topics are the  subject of a number  of papers, there
are very  few detailed  studies on the  photometric properties  of the
host galaxies of radio  galaxies.  Zirbel (1996a) analyzes the surface
brightness  profiles of  the  host galaxies  of low-redshift  powerful
radio  galaxies.   This  work  suggests  that  FR~Is  exhibit  a  weak
correlation between the  size of the host galaxy  and the radio power,
and  also  leads to  the  conclusion that  FR~I  and  FR~II inhabit  a
different population of host  galaxies. More specifically, while FR~Is
are typically associated with cD galaxies, FR~IIs are mostly associated
with a  subclass of elliptical  called ``N-galaxies''. In  ground based
images,  N-class  galaxies  appear  to have  peaked  radial  brightness
profiles, while in HST images it is clear that their peculiar profiles
are  explained  by  the  presence  of  a  bright  quasar-like  nucleus
surrounded by a  faint fuzz.  Therfore, when deriving  the host galaxy
profiles,  ground-based images  are unreliable  if a  powerful nuclear
source is  present.  

From an imaging  study of a sample of  radio-quiet quasars, radio-loud
quasars  and  radio galaxies,  Dunlop  et  al.   (2003) conclude  that
spheroidal  hosts  become   more  prevalent  with  increasing  nuclear
luminosity.  For nuclear luminosities $M_v < -23.5$, the hosts of both
radio-loud and radio-quiet quasars  are all massive ellipticals.  They
also   suggest  that  the   basic  properties   of  these   hosts  are
indistinguishable  from  those  of  quiescent,  evolved,  low-redshift
ellipticals of comparable mass.\\ The  aim of this paper is to provide
valuable infrared  photometric data of  a sample of 82  3CR radio galaxies
and to  compare these data with that obtained  at other wavelengths  for both
normal and  radio galaxies.  
To achieve this goal we use both ground-based and high angular resolution
HST  data to  properly  disentangle  the host  galaxy  from any  other
component  related to  e.g. the  AGN.  The  layout of  the paper  is as
follows.   In Section  2  we  summarize the  sample  selection and  in
Section 3  we give  details of the  observations, image  reduction and
point  spread function  determination.  In Section  4  we present  the
galaxy  luminosity profiles  and  profile fitting.   In  Section 5  we
discuss the results and in Section 6 we analize these results from the
point of  view of  a different classification  of the galaxies  of the
sample.   In  Section 7  the  black  hole  mass obtained  through  two
different methods is discussed. Finally, in Section 8 we summarize our
conclusions.\\


\section{Sample selection} 

The sample of galaxies we analyze belongs to the Revised Third Cambridge Catalogue (3CR) (Bennet 1962a, 1962b).
The 3CR catalog (Spinrad et al. 1985)
is selected at 178 MHz, it includes the brightest radio sources in the
northern  sky  ($S_{178} >10$  Jy)  and  contains  radio galaxies  and
quasars out  to redshifts of $\sim2$.  Since this  catalogue is selected
based only on the low-frequency radio properties of the sources, it is
an  excellent  database  to   study  the  morphology  and  photometric
attributes of  the objects without any  orientation bias.  We
chose  all  3CR  sources  with  $z<0.3$  with  the  primary  goal  of
characterizing the  radio galaxy hosts  mostly free from the  effects of
dust. 
There is a total  of 116 sources in the 3CR up  to a redshift  of
0.3, and we present here the photometric properties of 82 objects.
These objects were randomly observed as part of a HST snapshot program 10173.
Only 7 galaxies of the sample were observed as part of other programs. They are: 
3C 84 (P 7330), 3C 264 (P 7882), 3C 293 (P 9726), 3C 305 (P 7853), 3C 317 (P 7886), 
3C 338 (P 7453), and 3C 405 (P 7258).
Table 1 lists  the observed  galaxies together  with  the observation
date,  right  ascension  and  declination  for  epoch  2000,  
FR  class  (Fanaroff \& Riley, 1974) and morphological classification.
 
We used the morphological classification from NED (Nasa/Ipac  Extragalactic  Database)
and Zirbel (1996, Paper II). As can be noted from Table 1 many of the host galaxies
(57\%)  still remain  morphologically unclassified.   
Most  of  the  galaxies  for  which  a  classification  is
available in the literature are ellipticals.  Thirty four (72\%) out of
the 47 morphologically  classified   galaxies   are  reported   as
ellipticals,  while 7 (15\%) objects  are  classified as  early
spirals and 6 (13\%) as N type galaxies.\\

\section{Observations and data processing}
We use two different sets of near-infrared data: HST/NICMOS and 
Telescopio Nazionale
Galileo (TNG) data.  HST data are essential to derive the radial
brightness profile of galaxies located at $z > 0.1$ and particularly
for those  with a bright nucleus that outshines  the host galaxy
stellar emission.  TNG  images are crucial for nearby  galaxies ($z <
0.1$) which in general are  much larger than the  NIC-2 field of
view and  for which a  direct measure of  the background level  is not
feasible with the HST data alone.  In  the following  we  give  a  brief description  of  the
datasets and the data analysis procedures.\\

\subsection{HST Observations}
All objects in our sample were observed with NICMOS Camera 2 (NIC2) that has a field
of view of $19.2\arcsec \times 19.2\arcsec$ and a projected pixel size
of  $0.075\arcsec$. All images  were obtained  using the  F160W filter
(similar to the standard  $H$ band)  and have  the  same total
exposure  time of  1152 seconds. For only a few exceptions exposure times
were shorter. These are: 3C84, 3C264, and 3C317 for which exposure times
were 640, 448, and 640 seconds respectively.   
A  more detailed  technical description of  the observations
can  be found  in  Madrid et al. (2006) (hereafter M2006).\\
We  obtained  the HST/NICMOS  data from  the
Multimission  Archive  at  the  STScI  (MAST).   The  data  was
processed by standard OTFR (On The Fly Reprocesing) 
calibration pipeline.  We  performed  two corrections  subsequent to  the
calibration  pipeline: the  removal  of the  pedestal  effect and  the
masking of the coronographic spot.  Bad columns and bad pixels were masked
out too.   The pedestal effect  is stochastic noise  introduced during
the  detector  readout (Noll  et  al.,  2004). Each  of  the
quadrants of  the detector have a different offset. One in particular, appears darker
when compared to  the others.   We remove this  effect using the  IRAF task
{\sc  pedsub},  in  the  Space Telescope  Science  Analysis  System
(STSDAS).  NIC2  has a hole to allow  coronographic observations, this
hole produces  a spot on  our images and  therefore we mask  this spot
before image combination. Information is recovered using the 4-dither pattern 
used for the observations. We  combine the  four exposures
using the  Pyraf task {\sc multidrizzle} also in STSDAS (Koekemoer
\&  Fruchter,  2002).    The  final  image  is  the   output  of  {\sc
multidrizzle}, in units of  $e^-$ s$^{-1}$.  During our data reduction
we preserve both the original sky level and the original pixel size of
$0.075\arcsec$.\\ For the photometric  calibration of the reduced data
we used the  formula of Dickinson et al.  (2002)  to derive  a Vega-normalized
magnitude  for F160W  (NIC2).  We  then  have m(F160W)  = -2.5  log [PHOTFNU $\times$
CR/$f_\nu$(Vega)]\ = 22.107  - 2.5 log (CR),
where PHOTFNU = $1.49816 \times 10^{-6}$  Jy s DN$^{-1}$, and CR is  the count rate in
DN s$^{-1}$.  The  flux density of Vega in the F160W  band for NIC2 is
$f_\nu$(Vega) =  1043.5 Jy. These magnitude  units are different
from  the ST  instrumental magnitudes  used by  M2006.   We chose  the Vega  system
because it is more similar to the standard ground based near-IR system,
and for  the F160W filter  the difference between Vega-mag  and ST-mag
corresponds  to 3.68 mag.\\  A visual inspection  of  our final
images revealed that  the  amplifier glow  effect  residuals were  not
negligible. The  amplifier  glow is caused  by  the  readout
amplifier  situated close  to each  corner of  the detector,  and even
after  the correction  was applied  we noted  that residuals  for most
cases were around 10\% to 20\% above the background level. We trim the
border  of  the  images  to  obtain accurate  surface  photometry  and
eliminate the amplifier glow.  The  resulting final field of view
is $16.5\arcsec \times 16.5\arcsec$.\\

\subsection{TNG data}

We have used  the images taken by Marchesini et al. (2005). These images were obtained with the 3.6 m 
Telescopio Nazionale Galileo (TNG), the Italian national facility located at La Palma Island (Spain). These observations
were made in two runs on Jul. 8-12, 2000 and on Feb. 9-13, 2001. ARNICA (ARcetri Near Infrared CAmera) was used
in the first run and NICS (Near Infrared Camera Spectrometer) was used in the second run.
ARNICA uses a  $256 \times 256$ pixel NICMOS 3 array with a pixel size of $0.35\arcsec$ and a total field of view (FoV) 
of $1.5\arcmin \times 1.5\arcmin$ which is more than 5 times greater than the FoV of the HST NICMOS camera 2.
On the other hand, NICS uses a Rockwell $1024 \times 1024$ HgCdTe Hawaii array. NICS has two modes
available, the small field mode with a pixel size of $0.13\arcsec$ and a FoV of $2.2\arcmin \times 2.2\arcmin$,
and the large field mode in which the pixel size and FoV are doubled.
For both ARNICA and NICS observations the $K\arcmin$ filter was used. This filter has a central wavelength of 
$2.12\mu m$ and a $FWHM$ of $0.35 \mu m$. For NICS the total integration times of 20 min were used, while for ARNICA
total integration times ranged from 5 to 24 min depending on the source.  More detailed information about the individual
observations are listed in Table 3 of Marchesini et al. (2005).\\


\subsection{Background subtraction and PSF modeling}

Since suitable  fitting of radial  profiles at low  surface brightness
levels  is  a  delicate  task,   we  are  extremely  cautious  in  the
determination of  the background level and its  uncertainty.  First of
all, most of  the galaxies with $z<0.1$ cover  the whole field-of-view
of HST/NICMOS,  therefore the background cannot  be measured directly.
Furthermore, a detailed analysis of  the background level for the rest
of the  galaxies with smaller apparent diameter  (those with $z>0.1$),
shows  that  the background  is  not  constant.   In fact,  background
variations among these  images can be as large as  40\% or more, which
makes it impossible to remove it  using a common value for all images.
Therefore,  each image  of the  $z>0.1$ galaxies  are to  be carefully
examined in  order to  determine the proper background level.   However, our
analysis shows that, in each  image, a constant value across the whole
detector's  field-of-view  is  accurate   enough  to  remove  the 
background.   For  those images  where  this  task  was possible, 
background removal left residuals  not higher than 5\%.\\ For galaxies
with $z<0.1$ the  approach is completely different.  Since  we are not
able to determine the background on NICMOS images given that these
galaxies cover  the whole chip,  we use ground based  $K'$-band images
taken with  the TNG. Because of  the much larger field  of view, these
images allow  us to  easily determine the  sky level and  also to
obtain  accurate luminosity profiles  in the  external regions  of the
galaxies.  The next step is to simply determine the adequate offset in
order to  match the luminosity  profiles obtained from the  HST images
with  those  obtained  from the  TNG.   The  matching  was done  in  a
intermediate region of the luminosity profiles, avoiding the innermost
regions of the  TNG profiles, where seeing effects  are important, and
avoiding the outermost region of the HST profiles where background
effects are noticeable.  We find  that zero points needed to match HST
profiles  with  TNG  profiles   are  surprisingly  constant  over  the
observation dates and  the estimated errors were less  that 0.05 $mag$
$arcsec^{-2}$.  Finally,  we assume that the $H-K'$  color gradient is
null for our  galaxies and then we subtract the  adequate background level on
the  HST images  in  order to  match  the outermost  part  of the  HST
profiles with those  obtained from the TNG images.   The assumption of
null color gradient is supported by the fact that we do not detect any
$H-K'$ color gradient for the galaxies with $z>0.1$, where an accurate
background  determination on  both the TNG  and the HST  images is
possible.   Further  support  to  this  assumption is  found  in  M{\"
o}llenhoff \& Heidt (2001) (hereafter M\&H2001).  These authors showed
that  in  a  sample  of  40  bright  spiral  galaxies  the  structural
parameters for  both the bulge  and the disk components  are identical
for $JHK$ filters, within the error they estimate from the data.\\ PSF
modeling for  NIC2 is well  described by Suchkov  \& Krist (1998)
and references therein. PSF diffraction structures, such as Airy rings
and spikes,  grow larger with increasing wavelength  and therefore one
must be very cautious when using  infrared images. In our case, only a
few  cases  show a  bright  nuclear  source  which causes  bright  PSF
diffraction  patterns on the  images.  In  order to  get rid  of these
patterns we have used the  Tiny Tim modeling software (Krist, 1993) to
create suitable  PSF models  for each of  the selected  images.  These
models  were  then  subtracted   from  the  individual  images.   This
procedure works well  in most images,  if larger  residuals were
noticed    they   were   properly    masked   before    any   isophote
fitting. Besides,  during the fitting procedure we  also discarded the
inner 3  pixels ($0.23\arcsec$)  of the obtained  luminosity profiles.
This additional consideration avoids any fitting contamination related
to the PSF subtraction (see Section 4).


\section {Luminosity profiles and fitting functions}

There  are numerous  methods for  extracting luminosity  profiles. The
variety  includes  one-dimensional  analysis  (Bagget et  al.   1988),
ellipse  fitting to  the  isophotes of  the  galaxies (H\'eraudeau  \&
Simien 1996) and two-dimensional  analysis (Byun \& Freeman 1995).  At
first glance our  infrared images  show mostly  elliptical galaxies
with no trace of any further structures, and in a few cases (20\%) we
can  detect dust  lanes, jets  or  plumes.  Therefore,  we are  mostly
working with axisymmetric structures which can be easily analized with
azimuthal isophote fitting. For this purpose, we use the {\sc ellipse}
routine within STSDAS (Jedrzejewski  1987). Before extracting the
galaxy profile, we carefully mask  all spurious objects such as stars,
and residual  diffraction patterns.

In the case of two galaxies overlapping, once  we  obtain  the  luminosity profile  and  
structural  parameters (center   coordinates,  ellipticity,   and  position   angle   of  the
isophotes),  we  construct a  model  galaxy using {\sc  bmodel}
(within  STSDAS) that  is  subtracted from  the  original image.   The
resulting image is then used  to extract the luminosity profile of the
remaining  galaxy. This process  is repeated  several times  until the
luminosity profiles of both galaxies converge.\\ Isophote  fitting was  performed up  to  a count
level of 2 $\sigma_{sky}$ i.e.,  we stopped the fitting procedure when
the  isophote level is  around twice  the background  dispersion, which
corresponds  to  a  surface  magnitude  of  $m_{F160W}$  =  21.8  $mag$ $arcsec^{-2}$ 
on the HST/NICMOS images.  This  procedure is used to derive  the
luminosity profiles from both the TNG and the HST images.\\


\subsection{Profile fitting}

Similarly  to profile  extraction, luminosity  profile fitting  can be
done using a wide variety of functions.  The most common functions are
the classical de Vaucouleurs $R^{1/4}$ profile (de Vaucouleurs, 1948),
the exponential profile (Freeman, 1970), and the most general S\'ersic
profile  $R^{n}$ (S\'ersic, 1968).  Generally,  the  $R^{1/4}$ profile
describes  very  well the  luminosity  profile  of  bulges, while  the
exponential profile  is used to  fit the luminosity profile  of disks.
However, it was recently  found that the $R^{1/4}$ function does
not  describe  the  bulge  profile  for  many  ellipticals  correctly.
Therefore, many  authors use the  more general $R^{n}$ profile,  as is
the case  for Brown et al. (2003). However, for 45\% of our sample galaxies, we
find  that a  single  S\'ersic  profile still  cannot  fit the  data.
Therefore, we try  a more general case using  two S\'ersic profiles in
order to  fit the inner and  outer regions of  the luminosity profiles
respectively.  We find that the  outer profile regions are well fitted
with a S\'ersic  index $n \sim 1$, equivalent  to an exponential disk.
The adopted fitting function  has only 5  free parameters and  can be
written in the following form:
\begin{equation} \label{dV+exp}
I(r)=I_{s} exp\Big[-k\Big(\frac{r}r_{e}\Big)^{n}\Big]+I_{0} exp\Big(-\frac{r}r_{0}\Big)
\end{equation}
In  the above  equation the  first  term corresponds  to the  S\'ersic
profile, where $I_{s}$ is the intensity at $r = 0$, and $r_{e}$ is the
radius that encloses  half of the total luminosity  of the bulge (also
known as the  effective radius). For $n < 1$,  $k$ can be approximated
(with an  error smaller than 0.1\%) by  the relation $k =  2n - 0.324$
(Ciotti  1991).   The  second  term  corresponds  to  the  exponential
profile, where  $I_{0}$ is  the central intensity  and $r_{0}$  is the
length scale. However, as is pointed out by de Jong et al. (2004) the
exponential luminosity profile does not prove that these galaxies have real disks.
Kinematic data is required to confirm the presence of a disk in any
individual galaxy.\\
Figures  1a through 1s show the  luminosity profile for
all  the galaxies  of our  sample. In  these figures  NICMOS  data are
represented  by  squares  while  data  from  TNG  are  represented  by
triangles.   The fitting functions  are also  displayed for  the bulge
(short  dashed), disk  (long dashed)  and the  sum of  both components
(continuum line). The radius is in Kpc (horizontal bottom axis) and $arcsec$ 
(horizontal upper axis).\\  
Parameters described  in  the above  equation were  obtained
using the {\sc nfit} routine within STSDAS (Schombert \& Bothun 1987).
The fitting  procedure is  carried  out only  in the  $S/N >  2$
portion of  the galaxy surface  brightness profiles.  This is  done to
exlude the  regions in  the faint end  of the luminosity  profiles, in
which the error is large and  the information is poor. We also excluded 
the inner 3 pixels ($\sim0.20\arcsec$) of the luminosity profiles in order to avoid
any contamination resulting from the PSF subtraction. The errors on the parameters are
calculated following the method described by Coenda
et al.  (2005).  Briefly, this technique consists  in creating test
images  to which  we artificially  add and  subtract a  constant value
corresponding to  $\sigma_{sky}$.  We then extract  the new luminosity
profile as explained  above and we fit it  with Equation \ref{dV+exp}.
These newly obtained parameters give us the respective upper and lower
limits for the parameters.\\
Intensity  parameters  are  then  converted into  surface  brightness,
expressed in  $mag$ $arcsec^{-2}$ by the  equation $m=-2.5log(I)$, while
units of $r_e$,  and $r_0$ are converted to  Kpc.  Measured errors for
$r_e$ and $r_0$  are smaller than 15\% while for  $m_e$, and $m_0$ are
below 0.20 $mag$ $arcsec^{-2}$.   Total luminosities of both the bulge
and the exponential components  are finally computed using the derived
photometric  parameters  and  integrating  separately  both  terms  of
eq. 1 as follows:
\begin{equation}
\label{luminosidad}
L=\int_{0}^{\infty}I(r)2\pi rdr
\end{equation}
which yields:
\begin{equation}
L_{bulge}=I_{s}r_{e}^{2}\pi\frac{2}{n k^{2/n}} \Gamma(2/n) 
\end{equation}
for the bulge component, and
\begin{equation}
L_{disk}=2\pi I_{0}r_{0}^{2}
\end{equation}
for   the   exponetial   component.    $\Gamma(2/n)$  is   the   gamma
function. Note  that eq. 2 assumes that  the galaxy is
face  on.  Therefore,  the   intensities  in  eq. 1   were
corrected by inclination as  in Kent (1985). Total apparent magnitudes
were then  converted into absolute magnitudes.   Throughout this paper
we  assume a  Hubble  constant  $H_0$ =  71  $km$ $s^{-1}$  $Mpc^{-1}$
together with $\Omega_M$ = 0.27 and $\Omega_{\Lambda}$ = 0.73.\\


\section{Results and discussion} 

Table  2  lists  the   photometric  parameters  obtained  through  the
procedures described in the previous sections for all sample galaxies.
Columns  1,  2,  and 3  list  the  source  name, the  ellipticity  and
redshift,  respectively.  Bulge  parameters $m_e$,  $r_e$ and  $n$ are
listed in columns 4,  5, and  6, while columns  7 and 8  list disk
parameters $m_0$  (inclination corrected) and  $r_0$. Finally, columns
9, 10,  11, and 12 list  the bulge absolute  magnitudes, disk absolute
magnitudes, total  galaxy magnitudes and bulge to  disk ratio.\\ Forty
three  out of  82 galaxies  (52\%) show  a single  S\'ersic luminosity
profile  type.   The  remaining  39 galaxies  (48\%)  have  luminosity
profiles  that have  to be  fitted with  a combination  of  a S\'ersic
profile plus an  exponential profile. Among FR~Is, 10 out of 22 galaxies (45\%) show
single luminosity profiles, while 12 (55\%) show bulge+disk type profiles.
FR~IIs, show 33 out of 60 galaxies (55\%) with single S\'ersic luminosity profiles, 
and 27 galaxies (45\%) with bulge+disk type profiles.
However, we also have 4 interesting
cases:  3C28, 3C129.1,  3C184.1 and  3C388. These  galaxies show  $n >
0.75$. This is somewhat reminiscent of the so-called pseudo-bulges recently
discovered in some spiral galaxies.
Pseudobulges  are formed  through secular  evolution of  galaxies.  In
other words,  the bulge  is formed from  disk evolution,  allowing the
direct formation  of bulges from disks in  isolated galaxies (Kormendy \& Kennicutt 2004).
Unfortunately, we do not have enough information, such as velocity dispersions, to 
confirm this possibility.
3C348 is another interesting case since this galaxy has
a single  S\'ersic profile  with $n$  = 0.93 which  indicates it  is a
bulgeless galaxy. 
This result is unexpected  since, as was pointed out in  Section 2, most of
these galaxies are simply reported in the literature as ellipticals.
However, it is not clear whether or not the exponential profile is the
signature of  a disk in  all of them.   
To investigate if these are real disks, we have plotted in Fig.  2a-d
the distribution  of ellipticity $e$  for all galaxies of  the sample.
We  differenciate   between  galaxies  that  show   a  single  S\'ersic
luminosity  profiles   and  those  that   show  bulge+disk  luminosity
profiles, and also between galaxies classified as FR~I and FR~II.
Histograms are plotted in
bins  of  $\Delta  e  =0.05$.  Galaxies  that  show  a  single
luminosity  profile (panel  d) display  a distribution  in ellipticity
which is rather similar to  that displayed by  a  population of normal 
elliptical galaxies,  which has been shown  to peak at  $e \sim 0.15$
(Sandage, Freeman \& Stokes 1970;  Ryden 1992).  For such a sub-sample
we  obtain an average  ellipticity $e$  = 0.18$\pm$0.08.   The average
ellipticity for  galaxies that also show an exponential  component in their
luminosity  profiles   (panel  d  dashed   line)  is  instead   $e$  =
0.24$\pm$0.10. A Kolmogorov-Smirnov (K-S) test applied to these sub
samples indicates that  these  two data  sets  do not have different ellipticity distributions.  
We also note  that there are no differences between  FR~Is (panel b) and
FR~IIs  (panel  c) ellipticity  distributions.   We  obtain  $e$ =  0.21
$\pm0.11$ and 0.20 $\pm0.12$ for FR~I and FR~II galaxies respectively.\\
Figure  3a shows  the total  absolute magnitude  distribution  for all
galaxies of  the sample. Again, we have  differenciated between
FR~I (panel b)  and FR~II sources (panel c),  and also between galaxies
with a  single S\'ersic luminosity  profile (d). Total luminosity
distributions  for  all  the  classes  look similar.   
For the  whole  sample it peaks  at
$M_{total}$ =  -25.4 $\pm$0.7,  while for FR~I  and FR~II  sub samples
separately  these values  are $M_{total  FR~I}$ =  -25.6  $\pm$0.7 and
$M_{total  FR~II}$  =  -25.3  $\pm$0.6, respectively.   The  K-S  test
applied  to  FR~I and  FR~II  data  sets  indicates that  the  magnitude
distributions are not statistically different.

We  have  made a  similar  analysis  for  bulge and  disk  photometric
parameters for the whole sample as  well as for the FR~I and FR~II sub
samples.  The  average values for the  photometric parameters together
with their standard deviations are  listed in Table 3.  FR~I and FR~II
bulges  have   similar  properties  in  terms   of  their  photometric
parameters, $M_{bulge}$, $m_e$, and $r_e$.  The S\'ersic index average
value is $n$  = 0.40 $\pm$0.18 for FR~I galaxies,  while for the FR~II
sub  sample  it  is $n$  =  0.33  $\pm$0.08.  However, the K-S  test indicates that 
the distributions of the S\'ersic index for FR~I and FR~II are not statistically different.\\

FR~I and FR~II disks show some interesting differences.
As reported  in Table~3, the FR~Is' scale length ($r_0$) is on
average a factor of two larger  than that of FR~IIs, although the
spread in  $r_0$ is large,  thus they are consitent  within $1\sigma$.
Similarly,  the central surface  magnitude ($m_0$)  is 0.6 mag
dimmer in FR~I galaxies, but again consistent with that of FR~IIs within 
$1\sigma$.   Interestingly,  the  K-S  tests  applied  to  these  data
indicates that the $m_0$ and $r_0$ distributions are statistically different 
at 98\% and 99.9\% confidence level for FR~I and FR~II galaxies respectively.   
We also find that  the  mean total  disk  magnitudes  are  $M_{disk FR~I}$  =  -25.1
$\pm$0.6 and $M_{disk FR~II}$ =  -24.6 $\pm$0.6.  However, in this case
the  K-S  test  does  not  indicate that  the  the  distributions  are
statistically different.

Finally, the  bulge to disk luminosity  ratio is $b/d$  = 0.8$\pm$0.4
for FR~Is  while it is $b/d$  = 1.3$\pm$0.8 for FR~IIs. K-S  test 
does not indicate these data sets are statistically different.

\subsection{Correlation of structural parameters}
The  study of correlations  between photometric  parameters is  a good
tool  to understand  the structure  of galaxies  and  their evolution.
Many scaling relations for galaxies have been discovered this way,
such as the fundamental plane (Djorgovski \& Davis 1987, Dressler et al. 1987) for
elliptical   galaxies  and  the   Tully-Fisher  (1977)   relation  for
spirals.\\  The correlations  between structural  parameters  for both
bulges and disks obtained for our  sample are plotted in Fig. 4. Empty
triangles represent FR~IIs while filled squares represent FR~Is.  
Panels a and b show the effective surface magnitude ($m_e$)
and the  bulge total absolute magnitude  ($M_{bulge}$) plotted against
the  logarithm of  the effective  radius ($r_e$)  in Kpc.   From these
plots  we see  a clear  correlation between  these parameters,  in the
sense that large bulges have  a lower effective surface magnitudes and
higher  total luminosities.   The  first correlation  is the  Kormendy
relation  (Kormendy  1977) for  bulge  galaxies.  A linear  regression
applied to these data gives:\\
\begin{center}
$m_e$ = 16.6($\pm0.2$) + 3.2($\pm0.3$) log($r_e$/Kpc)\\
$M_{bulge}$ = -23.6($\pm0.3$) - 2.0($\pm0.4$) log($r_e$/Kpc)\\
\end{center}
The slope for the $m_e$ vs.  $r_e$ regression is quite similar to that
obtained by Veilleux  et al. (2006) for a sample  of 33 luminous, late
stage  galactic mergers and  also to  that obtained  by Dunlop  et al.
(2003) for  their sub  sample of  radio loud  quasars.  The  $m_e$ vs.
$r_e$ relationship for these type of galaxies shows a steeper slope to
that found for normal  ellipticals $\sim1.8$ (Pahre, 1999).\\ 
When  this analysis  is only applied to  FR~Is we obtain:
\begin{center}
$m_e$ = 16.7($\pm0.3$) + 2.5($\pm0.5$) log($r_e$/Kpc)\\
$M_{bulge}$ = -23.2($\pm0.3$) - 2.6($\pm0.5$) log($r_e$/Kpc)\\
\end{center}
While for FR~IIs we have:\\
\begin{center}
$m_e$ = 16.7($\pm0.2$) + 3.4($\pm0.4$) log($r_e$/Kpc)\\ 
$M_{bulge}$ = -23.7($\pm0.4$) - 1.9($\pm0.6$) log($r_e$/Kpc)\\
\end{center}
These results indicate that FR~Is and FR~IIs have similar Kormendy 
and $M_{bulge}$ vs. log($r_e$) relations, which is in agreement with
the result of Dunlop et al. (2003). These authors find similar Kormendy relations
for a sample of 13 radio-quiet quasars, 10 radio-loud  quasars and 10 radio galaxies.\\ 
We  also  investigate  the correlations for disks photometric  parameters.
Results are shown in Figs.  4c and 4d.  Disk parameters show the same trend as seen for
bulge parameters.  More explicitely, large disks have a lower central
surface  magnitude and  have higher  total luminosities.  However, in
this case the  data show a larger dispersion than that observed for the bulges.
Linear regression fits applied  to these data yield the following
results:\\
\begin{center}
$m_0$ = 16.5($\pm0.3$) + 2.0($\pm0.5$) log($r_0$/Kpc)\\
$M_{disk}$ = -22.8($\pm0.2$) - 2.6($\pm0.6$) log($r_0$/Kpc)\\
\end{center}
Intriguingly  , these  results are similar  to those  obtained by
M\&H (2001) for the disk components  of a sample of 40 bright spiral
galaxies. This result could give further  support to the idea that exponential
profiles in these galaxies are truly due to the presence of a disk. However,
as it is pointed out by de Jong et al. (2003), this exponential distribution
of light might not be in a disk configuration flattened by rotation as disks
of spiral galaxies are.
 

\section{FR~Is, LEGs, HEGs, and QSOs}
The  FR~I  and  FR~II  classification  is  only  based  on  the  radio
morphology of the source.  We then chose another classification scheme
which relies  on the spectroscopic  properties, and is more  likely to
reflect  the  physical  properties   of  the  central  AGN. In
particular,  we  adopt the  scheme  defined  by  Jackson \&  Rawlings
(1997). They  classify high  and low ionization  narrow-lined galaxies
(HEG  and LEG)  on  the basis  of  the equivalent  width  (EW) of  the
[OIII]5007 emission line and/or the [OII]/[OIII] ratio.  Galaxies with
[OII]/[OIII]  $>$  1 and/or  EW  $<$ $10\AA$  are  defined  as LEG  (Low
Excitation Galaxies).   Similarly, quasars (QSO) are  defined as those
sources for which at least one  broad line has been observed.
We find  in our sample 22 FR~Is (corresponding to 27\% of the sample), 
22 LEGs (27\%), 24 HEGs (29\%), and 13 QSOs (16\%). Only two FR~II objects
remain unclassified (3C 277.3 and 3C 346).
As we pointed out in Section 5, 55\% of FRI galaxies show bulge + exponential
type profiles, which is identical to the percentage obtained for LEGs.  
HEGs show less galaxies with bulge + exponential profiles (43\%), while QSOs
have only 31\% of the galaxies with the exponential component.  
It is interesting to note that even if  the FR~I  classification is  based on  
the radio morphology, it is known that most,  if not all, FR~I galaxies are weak
lined LEGs (Hine \& Longair, 1979; Laing, 1994).

Summarizing,  we divide our  objects into the  following spectral
classification scheme: FR~Is,  LEGs, HEGs and QSOs, and  we investigate the
statistical   properties    of   the   host    galaxies   under   this
classification.  Table~4 lists the photometric parameters for each of
these galaxy  classes.  All of them have bulges with  similar average
photometric properties,  and  even if  FR~Is  have  on  average greater  S\'ersic
indexes ($n$  = 0.40 vs. $n$ $\sim$ 0.34) the Kolmogorov - Smirnov test  applied to these
data  does not  indicate that  the $n$  distribution for  FR~I bulges  is
statistically different from that of LEGs, HEGs or QSOs.  

Differences between these classes arise when we compare disk photometric parameters.  
FR~Is have on average larger  disks ($r_0$ = 8.2 Kpc)  and dimmer central surface
magnitude ($m_0  = 18.1$) than those of LEGs ($m_0$ = 17.5, $r_0$ = 4.3), HEGs ($m_0$ = 17.5, $r_0$ = 3.6),
and QSOs ($m_0$ = 17.5, $r_0$ = 4.2). However, statistical  tests applied to the $m_0$ distributions show
that FR~Is differentiate only from HEGs and QSOs(99\% confidence level) and not from LEGs.
Same tests applied to $r_0$ data of these classes reveal that
FR~Is have a different distribution when compared with those of LEGs, HEGs and QSOs (99\% confidence level). 
In other words, these results are similar to those obtained in the previous section.
They suggest that, from a photometric point of view, FR~Is are slightly different from LEGs,
and these differences become stronger when FR~Is are compared to HEGs and QSOs.
We also ran statistical tests comparing FR~Is + LEGs to HEGs + QSOs. Results were similar to those
described above but with a slightly less statistical significance (98\% level). 
We find that FR~Is + LEGs have on average $m_0 = 17.8 \pm0.5$  $mag$ $arcsec^{-2}$ and 
$r_0 = 6.3 \pm2.0$ Kpc, while HEGs + QSOs have $m_0 = 17.5 \pm0.4$  $mag$ $arcsec^{-2}$ 
and $r_0 = 3.9 \pm1.2$ Kpc.


\section{Black hole mass vs. bulge near infrafred luminosity and S\'ersic index} 
 
Recent studies suggest that  central super  massive  black holes
(SMBH) reside  in all galaxies with a bulge, as a result  of past quasar
activity (Aller \& Rischstone 2002, and references therein). Moreover,
SMBH  are claimed  to be  related to  the host  galaxy  properties and
therefore  this implies  that SMBH  and galaxy  formation and evolution  are closely
correlated.   Kormendy \&  Richstone  (1995) showed  that  the mass of the SMBH is
correlated (with a considerable scatter) with both bulge luminosity and
bulge  mass.  This  result  was later  confirmed  and strengthened  by
Magorrian et al. (1998) for a sample of nearby galaxies with kinematic
data.\\  On the  other  hand, it  has  also been  shown that  $M_{SMBH}$
correlates  with the  S\'ersic  index  $n$ (Graham  et  al. 2003,  and
references  therein).  These  authors claim  that the  scatter  of the
$M_{SMBH}-n$  relation  is equivalent  to  the $M_{SMBH}-\sigma$  relation
(Ferrarese  \& Merrit,  2000;  Tremaine et  al  2002) and  it has  the
additional  advantage  of requiring  only  galaxy  images rather  than
spectra.\\ The superb  quality of HST/NICMOS  images, together
with  ground  based  information  for  nearby objects,  allows  us  to
determine $L_{bulge}$  and $n$ with  high accuracy.  Therefore  we can
estimate the black hole mass for  each galaxy of the sample using both
relations, and check whether the  results are consistent.  First of all we
use the  following relation with $L_{bulge}$. as taken  from Marconi \&
Hunt (2003):
\begin{center}
log($M_{SMBH}$) = 8.04 + 1.25 (log($L_{Hbulge}$) - 10.8)
\end{center}
where $M_{SMBH}$  is the mass of  the black hole  and $L_{H-bulge}$ is
the total  bulge luminosity  in the  H band in  solar units.   

Secondly,  we use the  Graham \& Driver (2006) relation:
\begin{center}
log($M_{SMBH}$) = 2.68 log(1/3$n$) + 7.82
\end{center}
where $n$ is the S\'ersic index as expressed in Equation \ref{dV+exp}.
We estimate  the mass of the central  black hole $M_{SMBH}$ using
the  values of  $M_{bulge}$ and  $n$ as  listed in  Table 2,  and the
results are reported  in Fig.  5.  In the  figure, empty triangles and
squares represent  FR~I and FR~II populations,  respectively. The data
show a large dispersion and the results obtained through these methods
differ on average by a factor of three.  We also used the quadratic $M_{SMBH} - n$ 
relation instead of the linear one given by Graham \& Driver (2006) and no
major differences were observed.
This result might reflect the fact that we do not  find any correlation between $M_{bulge}$ and $n$.
As can  be seen in  Fig.  6, neither  FR~I nor FR~II  bulge luminosities correlate with the S\'ersic index.
Note that similar bulge luminosity and $n$ distributions for FR~Is and FR~IIs
also imply similar $M_{SMBH}$ distribution for both sub samples.

 
\section {Conclusions}
We have analyzed the infrared photometric parameters of 82 galaxies belonging to the 3CR Catalogue. 
From the morphological point of view most of these galaxies are not yet
classified, and classified galaxies are mostly ellipticals. Nevertheless, we find that only 43 (52\%) of the galaxies 
have a single component luminosity profile.
The remaining 39 galaxies (48\%) need a second component with $n$ = 1 suggesting the idea that these 
galaxies have an exponential disk. However, we cannot confirm this possibility without additional kinematical information. 
When FR~Is and FR~IIs are differenciated we find the following results:\\
1) Forty five percent of FR~Is show single luminosity profile while for FR~IIs this percentage is 55\%. 
Reciprocally, 56\% of FR~I galaxies show bulge+exponential type profiles compared to 44\% of FR~II galaxies.\\
2) Analysis of the photometric parameters of both FR~I and FR~II galaxies indicates that they have similar
bulge and disk magnitudes. Moreover, both FR~I and FR~2 bulges follow similar Kormendy relations.
Even so , if we consider the exponential component we have that FR~Is have,
a) dimmer central surface magnitudes and, b) much larger scale lengths than FR~IIs. 
K-S tests applied to the $m_0$ and $r_0$ distributions confirm these results at $>98\%$ confidence level.\\

We also choose another classification scheme which relies on the spectroscopic properties, and is 
more likely to reflect the physical properties of the sources. 
We divided the sample galaxies in 4 groups, FR~Is, LEGs, HEGs and QSOs. The results of the analysis were 
similar to those obtained with the previous classification. FR~I galaxies show again for the exponential component  
dimmer central surface magnitudes and much larger scale lengths than the rest of the 
selected classes ($>98\%$ confidence level). However, we also find that these differences are less 
conspicuous when FR~Is are compared to LEGs.\\
Black hole masses were also calculated using two different methods, 
using the S\'ersic index and the bulge total luminosity. Results show a great dispersion and black hole 
masses obtained through $L_{bulge}$ are on average a factor of 3 higher than those obtained
through $n$. This result is not surprising since we did not find a clear correlation between 
$L_{bulge}$ and $n$.\\
Summarizing, past results on the host galaxies of FR~I and FR~II sources gave confusing results on 
their properties and morphologies. Thanks to the HST/NICMOS data we can now probe deeply 
into the nuclear regions  of the hosts and we can clearly distinguish the contributions from the host 
galaxies and the nuclear sources. We show that the host galaxies have very similar bulge properties 
while there seems to be a real difference in the properties or presence of disks. 
These results indicate that the formation histories of the different classes of radio galaxies may 
be significantly different.\\


\acknowledgments
Support for this work was provided by the National Science Foundation through grant N1183 from the 
Association of Universities for Research in Astronomy, Inc., under NSF cooperative agreement AST-0132798.\\
DM is supported by NASA LTSA NNG04GE12G.\\
This  research has  made  use  of the  NASA  Astrophysics Data  System Bibliographic services.  
This  research also made use made  use of the NASA/IPAC Extragalactic  Database (NED) which  is operated by  the Jet
Propulsion  Laboratory,  California  Institute  of  Technology,  under contract with  the National Aeronautics and  Space Administration.  



\clearpage

\begin{deluxetable}{lccccc}
\tabletypesize{\scriptsize}  
\tablecaption{Observation Log\label{tbl-1}}   
\tablewidth{0pt}
\tablehead{
\colhead{Source} & \colhead{Date (UT)} & \colhead{$\alpha$} & \colhead{$\delta$} & 
 \colhead{FR Class} & \colhead{Morph. Class.}\\
\colhead{(1)} & \colhead{(2)} & \colhead{(3)} & \colhead{(4)} & \colhead{(5)} & \colhead{(6)}
} 

\startdata

3C17      &  2006 Jul 2     &   00 38 20.5     &  -02 07 41.0      & II (QSO) & E pec \\
3C20      &  2005 Feb 27  &  00 43 09.27   &   +52 03 36.66   & II (HEG) & --- \\
3C28      &  2005 Jun 13  &  00 55 50.65   &   +26 24 36.93   & II (LEG) & E \\
3C31      &  2005 Jun 17  &  01 07 24.99   &   +32 24 45.02   & I & BCG; SA0- \\
3C33.1   &  2004 Aug 15 &  01 09 44.27   &   +73 11 57.2     & II (QSO) & ---   \\
3C35      &  2005 Mar 16  &  01 12 02.29   &   +49 28 35.33   &  II (LEG) & ---\\
3C52      &  2005 Mar 11  &  01 48 28.90   &   +53 32 27.9     & II (LEG) & --- \\
3C61.1   &  2004 Aug  9   &  02 22 36.00   &   +86 19 08.0     & II (LEG) & ---  \\
3C66B    &  2004 Nov  5   &  02 23 11.46   &   +42 59 31.34   & I & E \\
3C75N    &  2004 Nov 11 &  02 57 41.55   &   +06 01 36.58    & I & E0 \\
3C76.1    &  2005 Feb  6 &  03 03 15.0    &   +16 26 19.85      & I & E1? \\
3C79       &  2004 Oct 30 &  03 10 00.1    &   +17 05 58.91      & II (QSO) &  E \\
3C83.1    &  2005 Mar 12 &  03 18 15.8    &   +41 51 28.0        & I & E+  \\
3C84       & 1998 Mar 16  & 03 19 48        & +41 30 42             & I & cD pec\\
3C88       &  2004 Nov  6 &  03 27 54.17   &   +02 33 41.82      &  II (LEG) & E pec? \\
3C98       &  2005 Nov 25 &  03 58 54.4  &  +10 26 03              &  II (HEG) & E1?\\
3C105	&  2004 Oct 26 &  04 07 16.46   &   +03 42 25.68      & II (HEG) & E \\
3C111	&  2004 Dec  8 &  04 18 21.05   &   +38 01 35.77       &  II (QSO) & N \\
3C123	&  2004 Dec  7 &  04 37 04.4    &   +29 40 13.2          & II (LEG) & --- \\
3C129	 &  2004 Dec  8 &  04 49 09.07   &   +45 00 39.0         & I & E \\
3C129.1    &  2004 Nov 22 &  04 50 06.7    &   +45 03 06.0      & I & E \\
3C132       & 2005 Nov 24 &  04 56 43.0  &  +22 49 22              &II (LEG) & --- \\
3C133	 &  2004 Dec 13 &  05 02 58.4    &   +25 16 28.0         & II (LEG) & ---  \\
3C135	 &  2005 Apr  8 &  05 14 08.3    &   +00 56 32.0          & II (HEG) & E \\
3C153       &  2005 Nov 25 &  06 09 32.5  &  +48 04 15            &  II (LEG) & --- \\
3C165	 &  2005 Apr 26 &  06 43 06.6    &   +23 19 03.0         &  II (LEG) & ---  \\
3C166       &  2005 Nov 4 & 06 45 24.1  & +21 21 51                & II (LEG) & E \\
3C171	 &  2004 Nov 14 &  06 55 14.72   &   +54 08 58.27     & II (HEG)  & N  \\
3C173.1    &  2004 Nov 22 &  07 09 24.34   &   +74 49 15.19   & II (LEG)  & ---  \\
3C180	 &  2005 Feb 20 &  07 27 04.77   &   -02 04 30.97      & II (LEG) & --- \\
3C184.1    &  2004 Nov 26 &  07 43 01.28   &   +80 26 26.3     & II (QSO) & E \\
3C192	 &  2005 Jan  8 &  08 05 35.0    &   +24 09 50.0          & II (HEG) & ---  \\
3C196.1    &  2005 Feb  1 &  08 15 27.73   &   -03 08 26.99     & II (QSO)  & cD  \\
3C197.1    &  2005 Apr 19 &  08 21 33.7    &   +47 02 37.0       & II (HEG) & ---  \\
3C198	 &  2005 May  3 &  08 22 31.9    &   +05 57 7.0           & II (HEG) & E \\
3C213.1    &  2005 Feb 12 &  09 01 05.3    &   +29 01 46.0      & II (LEG) & --- \\
3C219	 &  2004 Sep 14 &  09 21 8.64    &   +45 38 56.49       & II (QSO) &  ---  \\
3C223	 &  2005 Feb 10 &  09 39 52.76   &   +35 53 59.12       & II (QSO) & E2 \\
3C223.1    &  2005 Jan 18 &  09 41 24.04   &   +39 44 42.39    &  II (HEG) & S0? \\
3C227	 &  2005 Mar 28 &  09 47 45.14   &   +07 25 20.33      & II (QSO) & N \\
3C234       &  2005 Nov 3    &10 01 49.5   &  +28 47 09            & II (QSO) & N \\
3C236	 &  2004 Nov  2   &  10 06 01.7    &   +34 54 10.0       & II (LEG)  & E \\
3C264  & 1998 May 12 & 11 45 05.0  & +19 36 23                    & I & E \\
3C277.3    &  2005 Mar 24  &  12 54 12.06   &   +27 37 32.66   & II & E  \\
3C284       &   2006 Mar 4  &  13 11 04.7  &  +27 28 08             & II (HEG) & --- \\
3C285	 &  2004 Dec  5 &  13 21 17.8    &   +42 35 15.0          & II (HEG) & --- \\
3C287.1    &  2005 Jul 16 &  13 32 53.27   &   +02 00 44.73     & II (QSO) & E pec \\
3C288	 &  2004 Oct 31 &  13 38 50.0    &   +38 51 10.7         & I & --- \\
3C293   & 2004 Mar 17 & 13 52 17.8 &  +31 26 46                   & II (LEG) & --- \\
3C296       &   2006 Apr 21 &  14 16 52.9  &  +10 48 26           & I & --- \\
3C300       &   2006 Mar 4 &  14 23 01.0  &  +19 35 17            & II (HEG) & E  \\
3C303	 &  2004 Dec 26 &  14 43 02.74   &   +52 01 37.5      & II (QSO) & N \\
3C305   & 1998 Jul 19 & 14 49 21.6 & +63 16 14                    & I & SB0 \\
3C310	 &  2004 Aug 13 &  15 04 57.18   &   +26 00 56.87    & I & ---  \\
3C314.1    &  2005 Feb 24 &  15 10 23.12   &   +70 45 53.4    & I & E \\
3C315	 &  2004 Dec 30 &  15 13 40.0    &   +26 07 27.0       & I & --- \\
3C317  &  1998 Aug 26  & 15 16 44.5 & +07 01 17                 & I & cD; E \\
3C319	 &  2004 Dec 29 &  15 24 05.5    &   +54 28 14.6       & II (LEG) & ---  \\
3C326       &   2006 Apr 21 &  15 52 09.1  &  +20 05 24          & II (LEG) & ---  \\
3C332       &  2006 Jan 12  &  16 17 42.5  &  +32 22 35          & II (QSO) & E \\
3C338      & 1997 Dec 17  &  16 28 38.5 & +39 33 06               & I & --- \\
3C346	 &  2005 May 19 &  16 43 48.69   &   +17 15 48.09     & II (LEG) & E  \\ 
3C348	 &  2005 May  9 &  16 51 08.16   &   +04 59 33.84      &  I & E \\
3C349	 &  2005 Mar 23 &  16 59 28.84   &   +47 02 56.8        & II (HEG) & --- \\
3C353	 &  2004 Sep  9 &  17 20 28.16   &   -00 58 47.06       & II (LEG) & SA0- \\ 
3C357       &  2006 Mar 25 &  17 28 18.5  &  +31 46 14            & II (HEG) & E  \\
3C379.1    &  2004 Nov  5 &  18 24 32.53   &   +74 20 58.64    &  II (HEG) & --- \\
3C381	 &  2004 Nov 11 &  18 33 46.29   &   +47 27 02.9        & II (HEG)  & ---  \\
3C386	 &  2005 Jun 15 &  18 38 26.27   &   +17 11 49.57      & I  & SA0-  \\
3C388	 &  2004 Oct 19 &  18 44 02.4    &   +45 33 30.0         & II (LEG) & --- \\
3C401	 &  2004 Aug 11 &  19 40 25.14   &   +60 41 36.85      & II (LEG) & E \\
3C402	 &  2004 Dec 10 &  19 41 46.0    &   +50 35 44.9          &  II (HEG) & S?\\
3C403	 &  2004 Nov  6 &  19 52 15.81   &   +02 30 24.4          & II (HEG) & S0 \\
3C405    &  1997 Dec 16 & 19 59 28.3 & +40 44 02                    & II (HEG) &  S?\\
3C424     &  2006 Jun 22 &  20 48 12.0    &   +07 01 17.0          & I & E \\
3C433	 &  2004 Aug 18 &  21 23 44.6    &   +25 04 28.5         & II (HEG)& ---  \\ 
3C436	 &  2004 Nov  9 &  21 44 11.74   &   +28 10 18.67       & II (HEG)  & --- \\
3C438	 &  2004 Nov 18 &  21 55 52.3    &   +38 00 30.0         & II (HEG)  & ---  \\
3C449	 &  2004 Nov 11 &  22 31 20.63   &   +39 21 30.07      & I & --- \\
3C452	 &  2004 Nov 28 &  22 45 48.9    &   +39 41 14.47      &  II (HEG) & E \\
3C459    &  2006 Jun 24  &  23 16 35.2    &   +04 05 18.0       & II (LEG) & N \\
3C465	 &  2004 Sep 28 &  23 38 29.41   &   +27 01 53.03     & I & cD; E+ pec \\

\enddata

\tablecomments{Col.   (1),  3CR  number;  col. (2)  HST observation  date;
col.  (3) right  ascension for  epoch 2000;  col. (4)  declination for
epoch 2000; col. (5) FR class; col (6) morphological classification.}

\end{deluxetable}


\clearpage

\begin{deluxetable}{lccccccccccc} 
\tabletypesize{\scriptsize} 
\tablecaption{Photometrical Properties of the NICMOS Snapshot Survey\label{tbl-2}} 
\tablewidth{0pt}
\tablehead{
\colhead{Source} & \colhead{ellip.} & \colhead{z} & \colhead{$m_e$} & 
\colhead{$r_e$} &  \colhead{$n$} & \colhead{$m_0$} & \colhead{$r_0$} & \colhead{$M_{bulge}$} & 
\colhead{$M_{disk}$} & \colhead{$M_{total}$} & \colhead{$b/d$} \\
\colhead{(1)} & \colhead{(2)} & \colhead{(3)} & \colhead{(4)} & \colhead{(5)} &
\colhead{(6)} & \colhead{(7)} & \colhead{(8)} & \colhead{(9)} & \colhead{(10)} & 
\colhead{(11)} & \colhead{(12)} 
} 

\startdata

3C17 &  0.25 &  0.21968 &  18.28 &   3.21 &  0.370 &     --- &     --- &  -24.88 &      --- &  -24.88 &     --- \\
3C20 &  0.05 &  0.17400 &  17.89 &   2.33 &  0.350 &     --- &     --- &  -24.43 &      --- &  -24.43 &     --- \\
3C28 &  0.18 &  0.19520 &  17.76 &   1.05 &  0.821 &  17.25 &   4.49 &  -22.49 &  -25.36 &  -25.43 &   0.07 \\
3C31 &  0.11 &  0.01670 &  18.09 &   5.22 &  0.272 &     --- &     --- &  -25.50 &      --- &  -25.50 &     --- \\
3C33.1 &  0.09 &  0.18090 &  18.65 &   3.08 &  0.443 &     --- &     --- &  -24.16 &      --- &  -24.16 &     --- \\
3C35 &  0.27 &  0.06700 &  18.77 &   5.89 &  0.371 &     --- &     --- &  -25.13 &      --- &  -25.13 &     --- \\
3C52 &  0.31 &  0.28540 &  19.06 &   8.34 &  0.240 &     --- &     --- &  -26.63 &      --- &  -26.63 &     --- \\
3C61.1 &  0.07 &  0.18400 &  19.52 &   2.86 &  0.282 &     --- &     --- &  -23.40 &      --- &  -23.40 &     --- \\
3C66.B &  0.15 &  0.02150 &  19.31 &   9.16 &  0.248 &  19.94 &  20.69 &  -25.58 &  -25.32 &  -26.21 &   1.27 \\
3C75.N &  0.03 &  0.02315 &  17.10 &   1.98 &  0.247 &     --- &     --- &  -24.41 &      --- &  -24.41 &     --- \\
3C76.1 &  0.15 &  0.03240 &  16.70 &   1.08 &  0.261 &  19.15 &   5.19 &  -23.45 &  -23.05 &  -24.02 &   1.46 \\
3C79 &  0.10 &  0.25595 &  18.38 &   4.27 &  0.348 &     --- &     --- &  -25.56 &      --- &  -25.56 &     --- \\
3C83.1 &  0.25 &  0.02550 &  18.93 &  12.01 &  0.214 &     --- &     --- &  -26.63 &      --- &  -26.63 &     --- \\
3C84   &  0.22   & 0.01756  & 18.73  &  6.52   & 0.356  & 18.22  &  8.12 &  -25.33  & -25.11 &  -25.98  &  1.23\\
3C88 &  0.35 &  0.03022 &  19.18 &   4.92 &  0.227 &  18.11 &   6.57 &  -24.42 &  -24.68 &  -25.31 &   0.79 \\
3C98 &  0.15 &  0.03000 &  18.31 &   3.12 &  0.248 &     --- &     --- &  -24.28 &      --- &  -24.28 &     --- \\
3C105 &  0.28 &  0.08900 &  17.16 &   1.19 &  0.315 &  17.55 &   2.20 &  -23.45 &  -23.12 &  -24.05 &   1.35 \\
3C111 &  0.25 &  0.04850 &  18.87 &   3.44 &  0.173 &     --- &     --- &  -24.16 &      --- &  -24.16 &     --- \\
3C123 &  0.12 &  0.21770 &  22.28 &  28.65 &  0.166 &     --- &     --- &  -26.04 &      --- &  -26.04 &     --- \\
3C129 &  0.13 &  0.02080 &  17.39 &   1.75 &  0.637 &  17.71 &   4.22 &  -23.33 &  -24.01 &  -24.47 &   0.54 \\
3C129.1 &  0.24 &  0.02220 &  17.32 &   1.76 &  0.900 &  17.76 &   5.97 &  -23.32 &  -24.78 &  -25.03 &   0.26 \\
3C132 &  0.18 &  0.21400 &  18.66 &   5.36 &  0.362 &     --- &     --- &  -25.60 &      --- &  -25.60 &     --- \\
3C133 &  0.06 &  0.27750 &  19.44 &   4.74 &  0.319 &     --- &     --- &  -24.82 &      --- &  -24.82 &     --- \\
3C135 &  0.19 &  0.12530 &  17.12 &   1.19 &  0.223 &  17.42 &   2.49 &  -23.78 &  -23.64 &  -24.47 &   1.14 \\
3C153 &  0.11 &  0.27700 &  16.79 &   1.28 &  0.259 &  18.44 &   5.27 &  -24.75 &  -24.81 &  -25.53 &   0.95 \\
3C165 &  0.18 &  0.29570 &  20.21 &   9.23 &  0.241 &     --- &     --- &  -25.72 &      --- &  -25.72 &     --- \\
3C166 &  0.07 &  0.24500 &  21.49 &  14.18 &  0.225 &     --- &     --- &  -25.24 &      --- &  -25.24 &     --- \\
3C171 &  0.08 &  0.23840 &  19.11 &   4.21 &  0.305 &     --- &     --- &  -24.78 &      --- &  -24.78 &     --- \\
3C173.1 &  0.23 &  0.29210 &  19.46 &   9.01 &  0.162 &     --- &     --- &  -26.62 &      --- &  -26.62 &     --- \\
3C180 &  0.43 &  0.22000 &  20.00  &   7.44 &  0.242 &  18.07 &   5.56 &  -25.21 &  -25.09 &  -25.91 &   1.11 \\
3C184.1 &  0.27 &  0.11820 &  15.79 &   0.59 &  0.904 &  16.74 &   2.05 &  -22.87 &  -23.88 &  -24.24 &   0.39 \\
3C192 &  0.03 &  0.05980 &  17.84 &   2.61 &  0.253 &     --- &     --- &  -24.39 &      --- &  -24.39 &     --- \\
3C196.1 &  0.29 &  0.19800 &  19.03 &   2.41 &  0.337 &  17.75 &   6.65 &  -23.47 &  -25.72 &  -25.85 &   0.13 \\
3C197.1 &  0.08 &  0.13010 &  19.42 &   4.73 &  0.227 &     --- &     --- &  -24.51 &      --- &  -24.51 &     --- \\
3C198 &  0.15 &  0.08150 &  19.31 &   3.50 &  0.292 &     --- &     --- &  -23.61 &      --- &  -23.61 &     --- \\
3C213.1 &  0.39 &  0.19400 &  18.95 &   4.59 &  0.281 &     --- &     --- &  -25.01 &      --- &  -25.01 &     --- \\
3C219 &  0.23 &  0.17440 &  18.92 &   6.27 &  0.399 &     --- &     --- &  -25.48 &      --- &  -25.48 &     --- \\
3C223 &  0.18 &  0.13680 &  18.71 &   4.06 &  0.286 &     --- &     --- &  -24.76 &      --- &  -24.76 &     --- \\
3C223.1 &  0.45 &  0.10700 &  16.99 &   1.50 &  0.261 &  16.74 &   3.21 &  -24.25 &  -24.78 &  -25.30 &   0.61 \\
3C227 &  0.17 &  0.08610 &  18.47 &   2.93 &  0.235 &     --- &     --- &  -24.21 &      --- &  -24.21 &     --- \\
3C234 &  0.07 &  0.18500 &  17.88 &   2.78 &  0.433 &     --- &     --- &  -24.76 &      --- &  -24.76 &     --- \\
3C236 &  0.34 &  0.10050 &  19.57 &  10.06 &  0.291 &  16.62 &   1.91 &  -25.72 &  -23.75 &  -25.89 &   6.15 \\
3C264  &  0.05  & 0.02172 & 16.13  & 1.52   & 0.671  & 17.70  & 5.48  & -24.37  & -24.70   & -25.30  & 0.74 \\
3C277.3 &  0.05 &  0.08570 &  19.11 &   5.54 &  0.255 &     --- &     --- &  -24.93 &      --- &  -24.93 &     --- \\
3C284 &  0.03 &  0.23900 &  19.44 &   6.13 &  0.213 &  17.59 &   2.00 &  -25.48 &  -23.42 &  -25.63 &   6.69 \\
3C285 &  0.42 &  0.07940 &  18.63 &   3.00 &  0.239 &  17.87 &   4.89 &  -24.08 &  -24.49 &  -25.05 &   0.69 \\
3C287.1 &  0.12 &  0.21590 &  16.24 &   1.00 &  0.268 &  18.26 &   4.83 &  -24.55 &  -24.58 &  -25.32 &   0.97 \\
3C288 &  0.03 &  0.24600 &  18.94 &   4.79 &  0.749 &  18.12 &   7.73 &  -24.81 &  -25.83 &  -26.19 &   0.39 \\
3C293 & 0.55  & 0.04503   & 17.44  &  1.63 &  0.968  & 17.43  &  4.65  & -23.09 &  -24.66  & -24.89  &  0.23\\
3C296 &  0.21 &  0.02370 &  17.97 &   6.60 &  0.277  &    ---  &   ---  &  -26.17 &   ---  &  -26.17  &   --- \\
3C300 &  0.29 &  0.27000 &  19.31 &   4.92 &  0.299 &     --- &     --- &  -24.92 &      --- &  -24.92 &     --- \\
3C303 &  0.10 &  0.14100 &  18.93 &   5.30 &  0.259 &     --- &     --- &  -25.19 &      --- &  -25.19 &     --- \\
3C305 &  0.35  & 0.04164  & 17.39  &  2.58  & 0.224  & 17.46 &   5.44 &  -24.88 &  -24.97 &   -25.67  &  0.92\\
3C310 &  0.15 &  0.05350 &  18.09 &   2.56 &  0.285 &  18.85 &   8.40 &  -24.06 &  -24.55 &  -25.09 &   0.63 \\
3C314.1 &  0.41 &  0.11970 &  18.99 &   4.44 &  0.408 &     --- &     --- &  -24.43 &      --- &  -24.43 &     --- \\
3C315 &  0.35 &  0.10830 &  16.83 &   1.39 &  0.252 &  18.84 &   4.59 &  -24.28 &  -23.47 &  -24.70 &   2.11 \\
3C317 & 0.30  & 0.03446  & 17.83  &  1.63  & 0.434  & 17.24  &  7.63 &  -23.06 &  -25.89 &  -25.97 &   0.07\\
3C319 &  0.24 &  0.19200 &  16.40 &   0.78 &  0.353 &  18.61 &   4.05 &  -23.59 &  -23.74 &  -24.42 &   0.87 \\
3C326 &  0.39 &  0.08900 &  17.13 &   1.38 &  0.269 &  16.90 &   2.07 &  -23.86 &  -23.62 &  -24.50 &   1.25 \\
3C332 &  0.05 &  0.15150 &  15.94 &   0.88 &  0.254 &  17.82 &   3.10 &  -24.35 &  -23.82 &  -24.87 &   1.63 \\
3C338 &  0.28 &  0.03035  & 19.03 &  12.86 &  0.474 &     --- &     ---  & -26.29  &     ---  & -26.29  &    ---\\
3C346 &  0.30 &  0.16100 &  14.90 &   0.46 &  0.270 &  17.83 &   3.82 &  -23.98 &  -24.29 &  -24.90 &   0.76 \\
3C348 &  0.22 &  0.15400 &  19.58 &   9.25 &  0.925 &     --- &     --- &  -25.16 &      --- &  -25.16 &     --- \\
3C349 &  0.48 &  0.20500 &  17.27 &   1.48 &  0.249 &  17.89 &   3.45 &  -24.20 &  -24.02 &  -24.87 &   1.18 \\
3C353 &  0.03 &  0.03043 &  17.86 &   1.85 &  0.251 &  18.11 &   3.43 &  -23.53 &  -23.24 &  -24.15 &   1.31 \\
3C357 &  0.32 &  0.16700 &  18.94 &   5.29 &  0.238 &  17.94 &   5.17 &  -25.34 &  -24.87 &  -25.88 &   1.54 \\
3C379.1 &  0.11 &  0.25600 &  18.14 &   2.77 &  0.225 &  17.83 &   5.02 &  -25.06 &  -25.22 &  -25.89 &   0.87 \\
3C381 &  0.17 &  0.16050 &  17.10 &   1.56 &  0.264 &  17.78 &   2.86 &  -24.43 &  -23.71 &  -24.88 &   1.95 \\
3C386 &  0.12 &  0.01700 &  18.32 &   3.38 &  0.226 &     --- &     --- &  -24.38 &      --- &  -24.38 &     --- \\
3C388 &  0.13 &  0.09100 &  17.64 &   2.21 &  1.227 &  16.86 &   5.37 &  -23.62 &  -25.71 &  -25.86 &   0.14 \\
3C401 &  0.18 &  0.20104 &  17.89 &   1.17 &  0.310 &  17.97 &   5.34 &  -22.95 &  -24.89 &  -25.06 &   0.17 \\
3C402 &  0.21 &  0.02390 &  18.85 &   5.00 &  0.190 &     --- &     --- &  -24.76 &      --- &  -24.76 &     --- \\
3C403 &  0.26 &  0.05900 &  18.01 &   4.72 &  0.405 &     --- &     --- &  -25.32 &      --- &  -25.32 &     --- \\
3C405  & 0.25  & 0.05608  & 20.01 &  14.85 &  0.525 &     --- &     --- &  -25.67 &      --- &  -25.67 &     ---\\
3C424 &  0.03 &  0.12699 &  19.38 &   3.67 &  0.254 &     --- &     --- &  -23.91 &      --- &  -23.91 &     --- \\
3C433 &  0.43 &  0.10160 &  17.90 &   2.51 &  0.494 &  16.85 &   4.68 &  -24.08 &  -25.44 &  -25.71 &   0.29 \\
3C436 &  0.18 &  0.21450 &  19.12 &   6.61 &  0.255 &     --- &     --- &  -25.75 &      --- &  -25.75 &     --- \\
3C438 &  0.10 &  0.29000 &  19.86 &   9.96 &  0.298 &     --- &     --- &  -26.06 &      --- &  -26.06 &     --- \\
3C449 &  0.48 &  0.01710 &  19.01 &   5.08 &  0.164 &     --- &     --- &  -24.75 &      --- &  -24.75 &     --- \\
3C452 &  0.24 &  0.08110 &  18.15 &   3.83 &  0.276 &     --- &     --- &  -25.00 &      --- &  -25.00 &     --- \\
3C459 &  0.15 &  0.21990 &  15.45 &   0.81 &  0.430 &  17.59 &   3.38 &  -24.63 &  -24.48 &  -25.31 &   1.14 \\
3C465 &  0.20 &  0.03030 &  17.67 &   3.92 &  0.311 &  18.52 &  14.59 &  -25.25 &  -25.98 &  -26.43 &   0.51 \\

\enddata

\tablecomments{Col.  (1),  3CR number;  col (2), ellipticity;  col. (3),
redshift; col.  (4), effective surface magnitude; col.  (5), effective radius; col. (6), S\'ersic index; col. (7), central surface magnitude;  
col.  (8), scale length; col. (9), bulge absolute magnitude; col. (10), disk absolute magnitude; col. (11), total absolute magnitude; 
col. (12), bulge/disk ratio.}

\end{deluxetable}


\clearpage

\begin{deluxetable}{cccccccccc}  
\tabletypesize{\scriptsize}
\tablecaption{Average parameters for FR~I and FR~II galaxies\label{tbl-3}}
\tablewidth{0pt}
\tablehead{
\colhead{Galaxies} & \colhead{$M_{total}$} & \colhead{$M_{bulge}$} & \colhead{$M_{disk}$} &  
\colhead{$m_e$} & \colhead{$r_e$} & \colhead{$n$} & \colhead{$m_0$} & \colhead{$r_0$} & \colhead{B/D}\\
\colhead{} & \colhead{} & \colhead{} & \colhead{} & \colhead{$[mag/arcsec^2]$} & 
\colhead{[Kpc]} & \colhead{} & \colhead{$[mag/arcsec^2]$} & \colhead{[Kpc]} & \colhead{ }\\
\colhead{(1)} & \colhead{(2)} & \colhead{(3)} & \colhead{(4)} & \colhead{(5)} &
\colhead{(6)} & \colhead{(7)} & \colhead{(8)} & \colhead{(9)} & \colhead{(10)} 
}

\startdata

ALL   & -25.4 $\pm$ 0.7 & -25.0 $\pm$ 0.9 & -24.8 $\pm$ 0.6 & 17.6 $\pm$ 1.1 & 4.6 $\pm$ 2.7 & 0.35 $\pm$ 0.18 & 17.6 $\pm$ 0.7 & 5.3 $\pm$ 2.8 & 1.1 $\pm$ 1.0\\
FR~I  & -25.6 $\pm$ 0.7 & -25.1 $\pm$ 0.9 & -25.1 $\pm$ 0.6 & 17.7 $\pm$ 0.8 & 4.7 $\pm$ 2.6 & 0.40 $\pm$ 0.18 & 18.1 $\pm$ 0.7 & 8.2 $\pm$ 2.7 & 0.8 $\pm$ 0.4\\
FR~II & -25.3 $\pm$ 0.6 & -25.0 $\pm$ 0.7 & -24.6 $\pm$ 0.6 & 17.5 $\pm$ 1.0 & 4.6 $\pm$ 2.6 & 0.33 $\pm$ 0.08 & 17.5 $\pm$ 0.4 & 4.0 $\pm$ 1.3 & 1.3 $\pm$ 0.8\\

\enddata

\tablecomments{Col.  (1),  Sub sample galaxies; (2),  total absolute magnitude; Col. (3),
 bulge total magnitude;  col. (4), disk total magnitude; col.  (5), bulge effective magnitude; col.  (6), 
bulge effective radius ; col.  (7), S\'ersic index; col. (8), disk central surface magnitude; col. (9), disk scale length; col. (10), bulge to disk ratio.}

\end{deluxetable}


\clearpage

\begin{deluxetable}{cccccccccc}  
\tabletypesize{\scriptsize}
\tablecaption{Average parameters for FR~I, LEG, HEG and QSO galaxies\label{tbl-4}}
\tablewidth{0pt}
\tablehead{
\colhead{Galaxies} & \colhead{$M_{total}$} & \colhead{$M_{bulge}$} & \colhead{$M_{disk}$} &  
\colhead{$m_e$} & \colhead{$r_e$} & \colhead{$n$} & \colhead{$m_0$} & \colhead{$r_0$} & \colhead{B/D}\\
\colhead{} & \colhead{} & \colhead{} & \colhead{} & \colhead{$[mag/arcsec^2]$} & 
\colhead{[Kpc]} & \colhead{} & \colhead{$[mag/arcsec^2]$} & \colhead{[Kpc]} & \colhead{ }\\
\colhead{(1)} & \colhead{(2)} & \colhead{(3)} & \colhead{(4)} & \colhead{(5)} &
\colhead{(6)} & \colhead{(7)} & \colhead{(8)} & \colhead{(9)} & \colhead{(10)}
}

\startdata

FR~I & -25.6 $\pm$ 0.7 & -25.1 $\pm$ 0.7 & -25.1 $\pm$ 0.7 & 17.7 $\pm$ 0.9 & 4.7 $\pm$ 2.6 & 0.40 $\pm$ 0.16 & 18.1 $\pm$ 0.6 & 8.2 $\pm$ 3.1 & 0.8 $\pm$ 0.4\\
LEG  & -25.5 $\pm$ 0.6 & -25.2 $\pm$ 0.9 & -24.7 $\pm$ 0.6 & 17.7 $\pm$ 1.4 & 5.8 $\pm$ 4.1 & 0.38 $\pm$ 0.10 & 17.5 $\pm$ 0.5 & 4.3 $\pm$ 1.1 & 1.2 $\pm$ 0.8\\
HEG  & -25.2 $\pm$ 0.5 & -24.9 $\pm$ 0.6 & -24.5 $\pm$ 0.7 & 18.1 $\pm$ 0.8 & 4.2 $\pm$ 2.0 & 0.29 $\pm$ 0.10 & 17.5 $\pm$ 0.3 & 3.6 $\pm$ 1.1 & 1.6 $\pm$ 0.9\\
QSO  & -25.0 $\pm$ 0.5 & -24.7 $\pm$ 0.6 & -24.8 $\pm$ 0.8 & 17.3 $\pm$ 0.7 & 3.1 $\pm$ 1.2 & 0.36 $\pm$ 0.11 & 17.5 $\pm$ 0.4 & 4.2 $\pm$ 1.8 & 0.8 $\pm$ 0.5\\

\enddata

\tablecomments{Col.  (1),  Sub sample galaxies; (2),  total absolute magnitude; Col. (3),
 bulge total magnitude;  col. (4), disk total magnitude; col.  (5), bulge effective magnitude; col.  (6), 
bulge effective radius ; col.  (7), S\'ersic index; col. (8), disk central surface magnitude; col. (9), disk scale length; col. (10), bulge to disk ratio.}

\end{deluxetable}


\newcounter{subfig}

\clearpage

\renewcommand{\thefigure}{\arabic{figure}\alph{subfig}}
\setcounter{subfig}{1}

\begin{figure}
\plotone{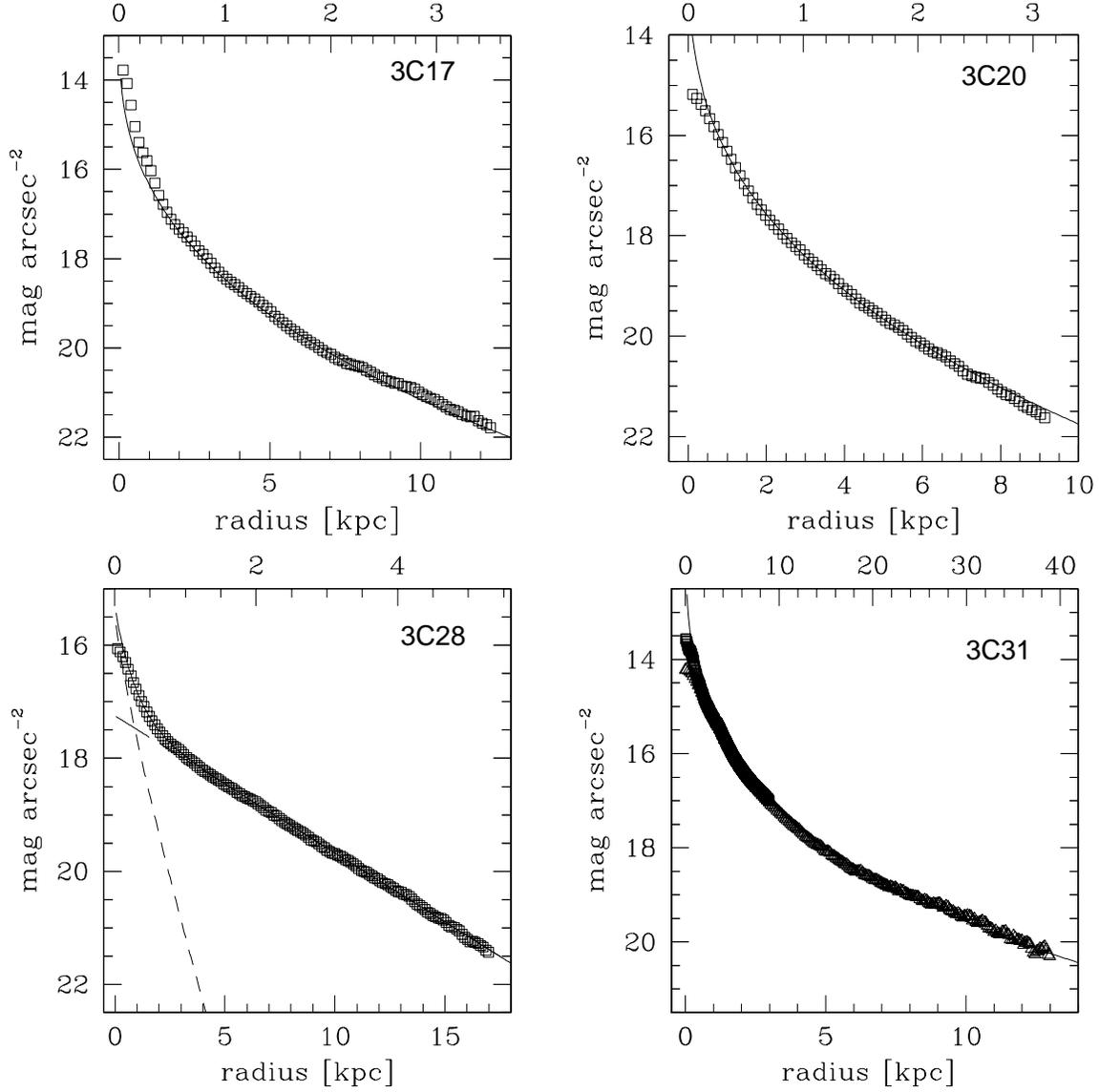}
\caption{Luminosity profiles. Squares represent NICMOS data while triangles represent TNG data. Upper ordinates give radius in units
of $arcsec$. Solid line shows the sum of the bulge (short dashed) plus disk (long dashed) fitting functions.}
\end{figure}


\clearpage

\addtocounter{figure}{-1}
\addtocounter{subfig}{1}

\begin{figure}
\plotone{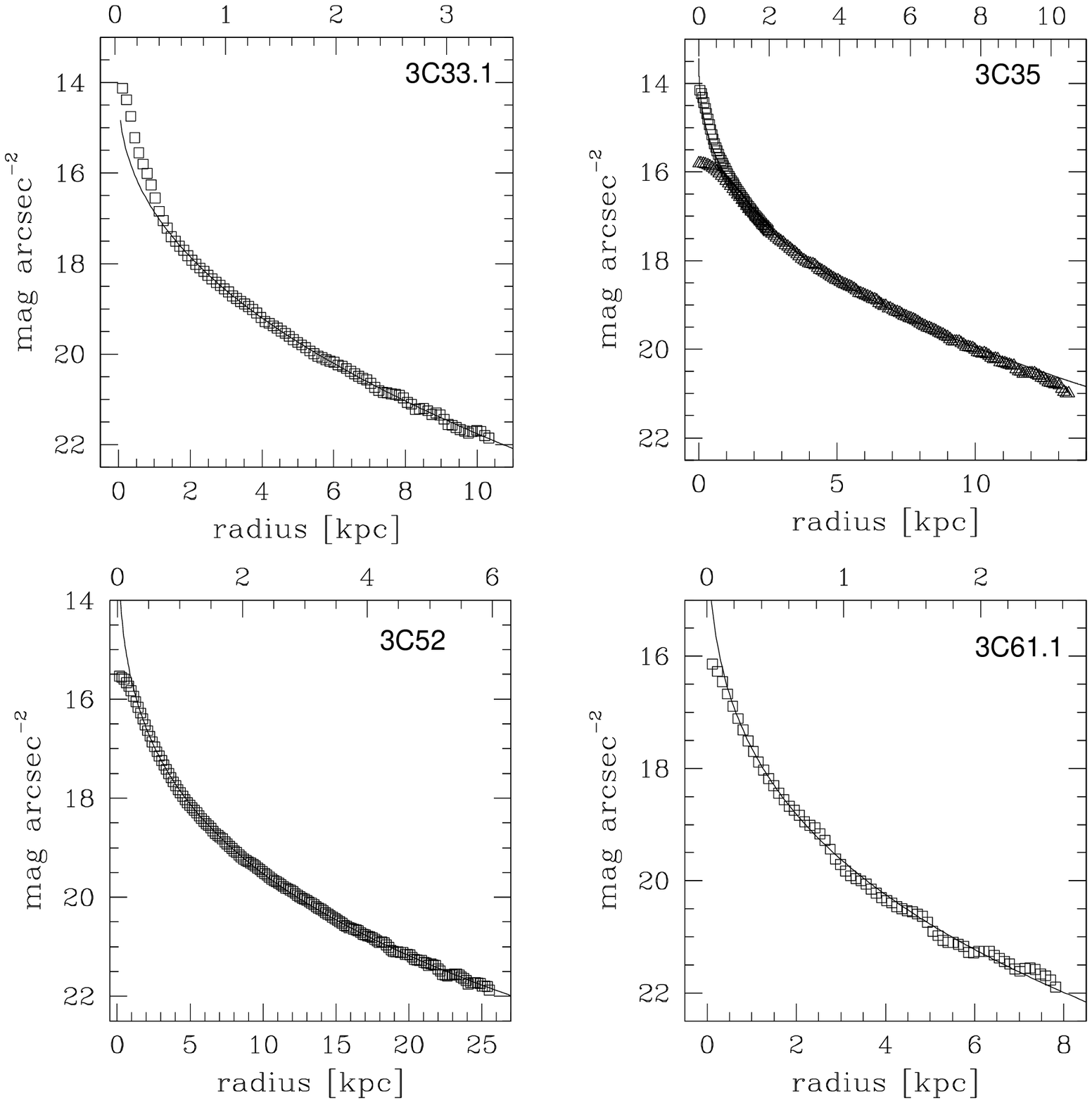}
\caption{Same as Fig. 1a.}
\end{figure}


\clearpage

\addtocounter{figure}{-1}
\addtocounter{subfig}{1}

\begin{figure}
\plotone{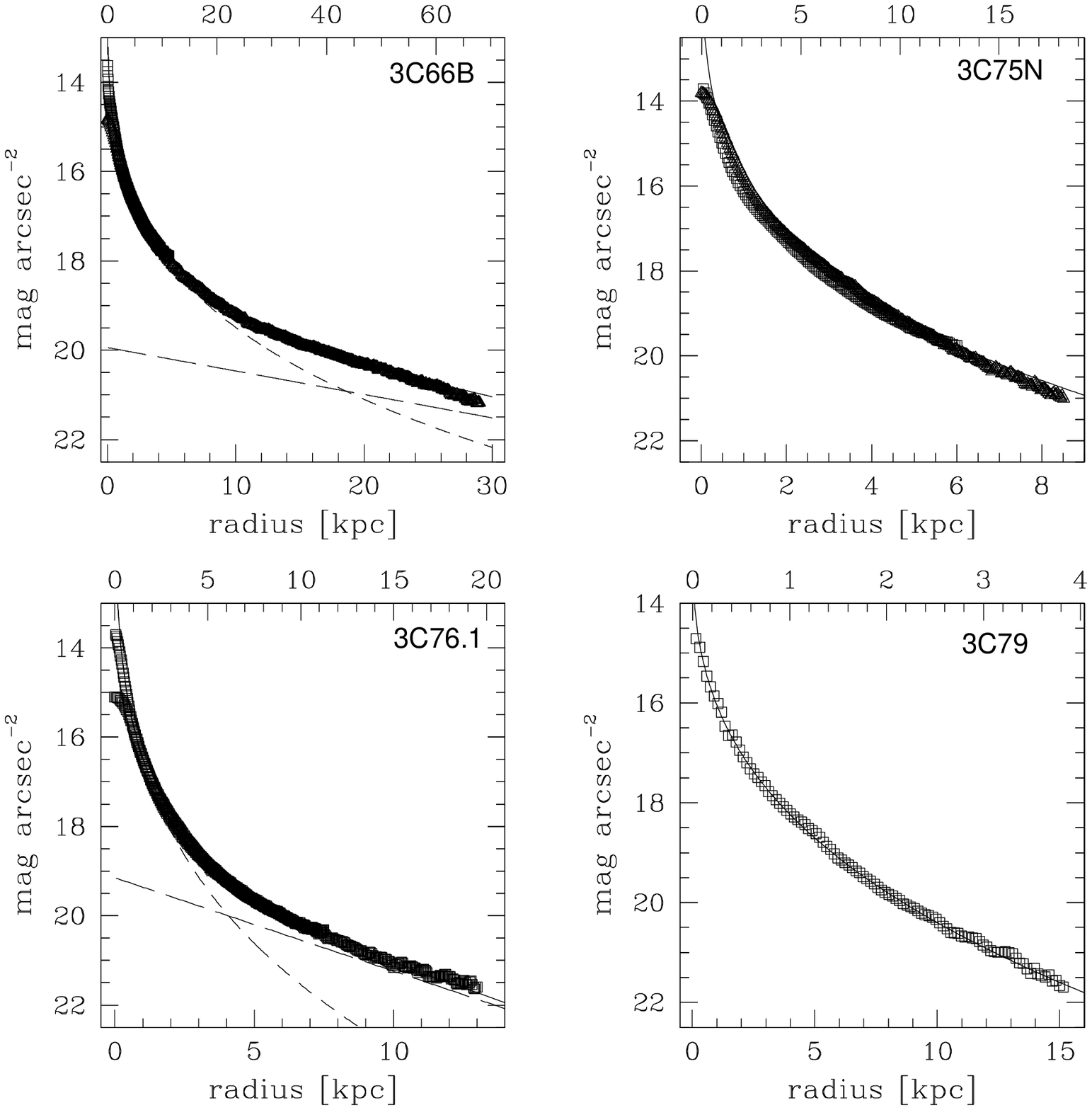}
\caption{Same as Fig. 1a.}
\end{figure}


\clearpage
\addtocounter{figure}{-1}
\addtocounter{subfig}{1}

\begin{figure}
\plotone{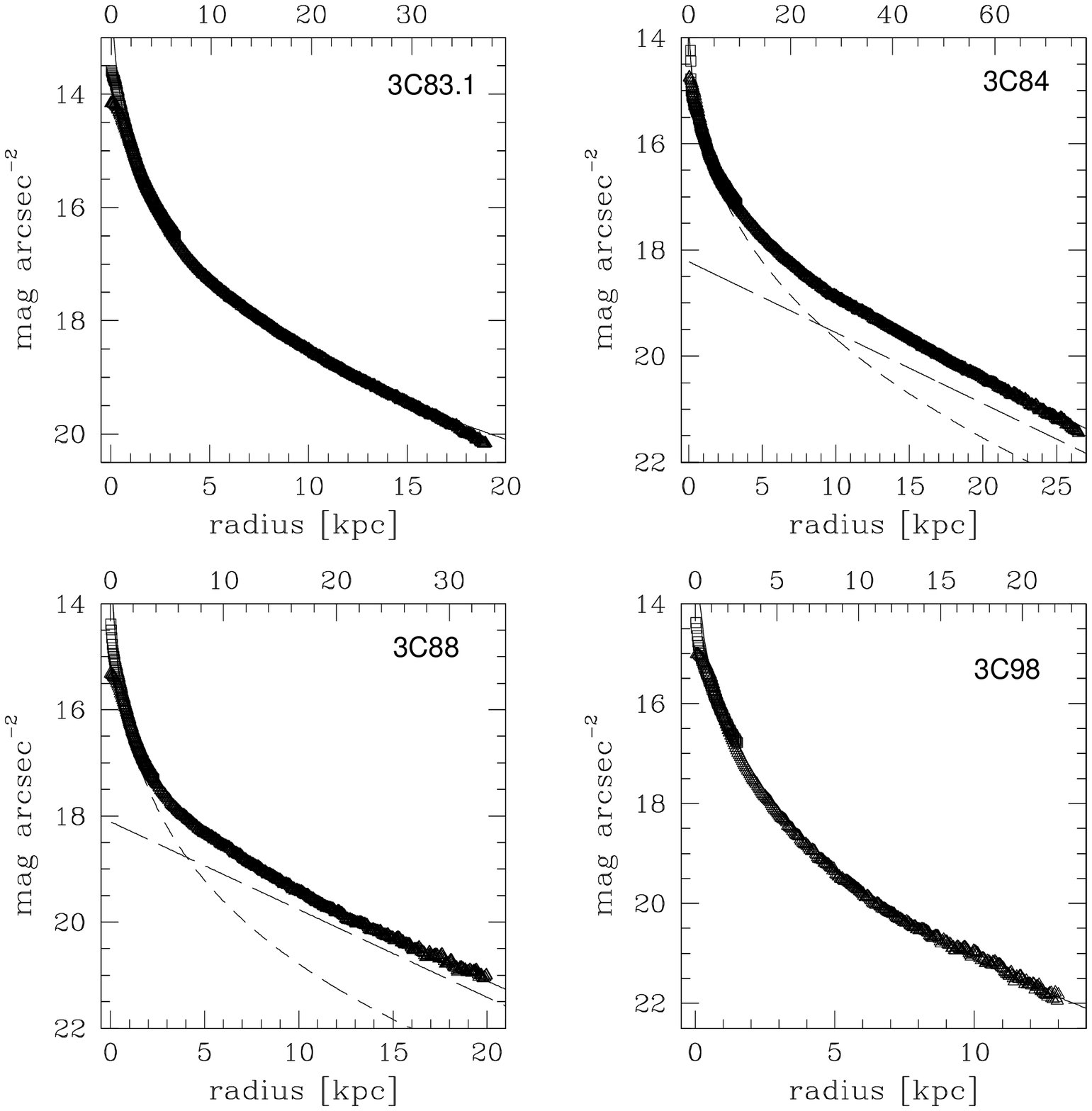}
\caption{Same as Fig. 1a.}
\end{figure}


\clearpage

\addtocounter{figure}{-1}
\addtocounter{subfig}{1}

\begin{figure}
\plotone{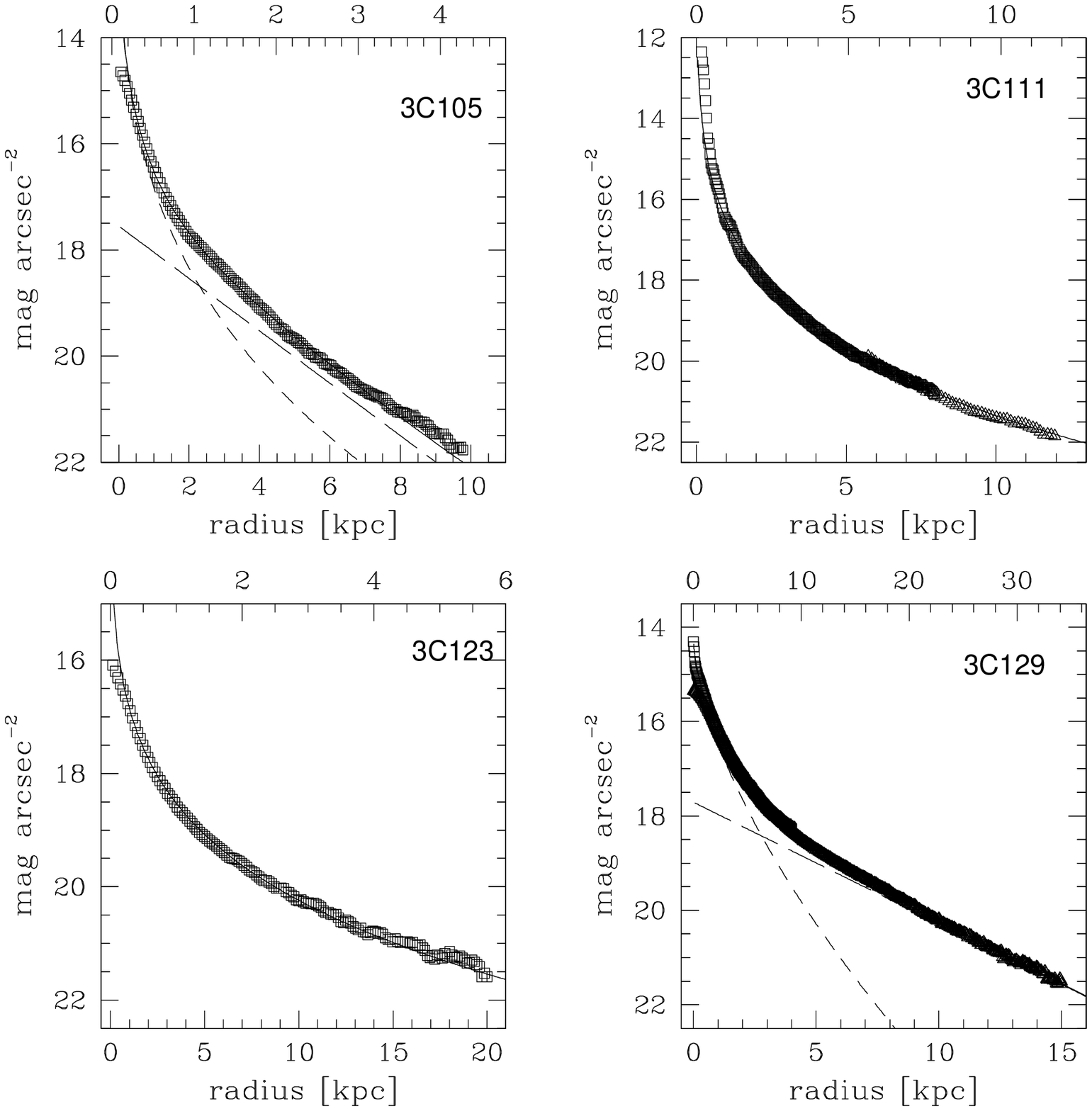}
\caption{Same as Fig. 1a.}
\end{figure}

\clearpage

\addtocounter{figure}{-1}
\addtocounter{subfig}{1}

\begin{figure}
\plotone{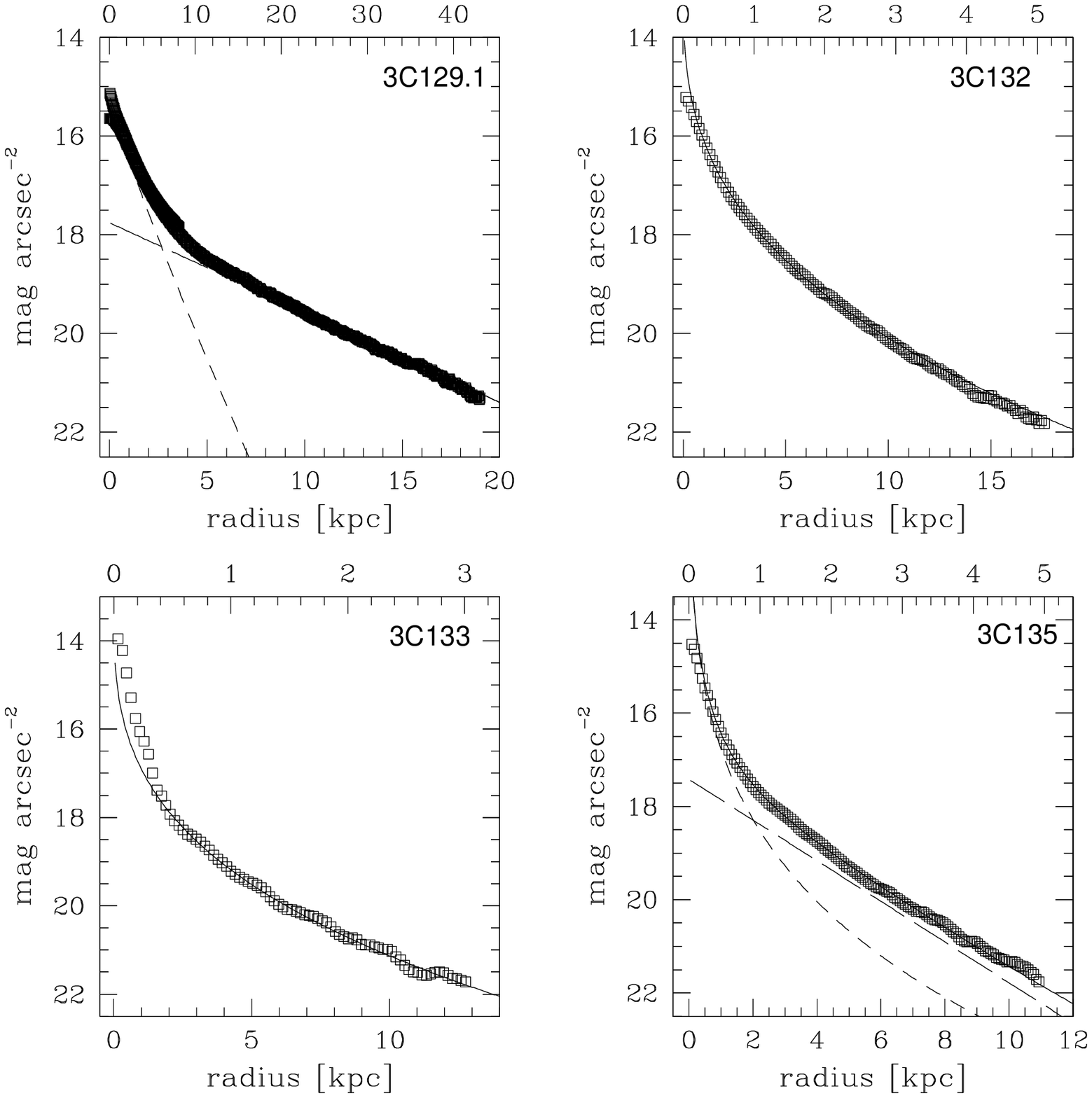}
\caption{Same as Fig. 1a.}
\end{figure}

\clearpage

\addtocounter{figure}{-1}
\addtocounter{subfig}{1}

\begin{figure}
\plotone{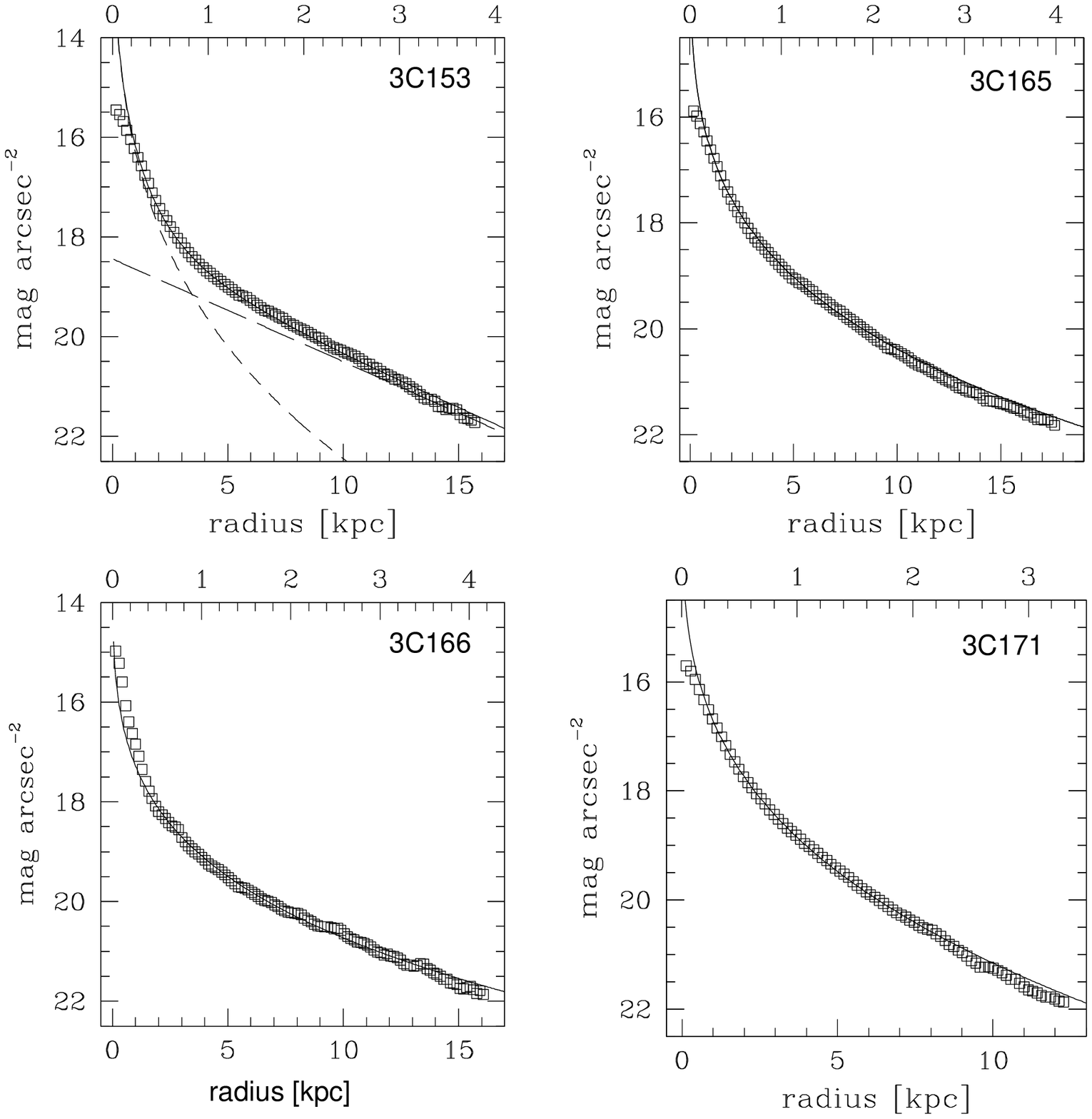}
\caption{Same as Fig. 1a.}
\end{figure}

\clearpage

\addtocounter{figure}{-1}
\addtocounter{subfig}{1}

\begin{figure}
\plotone{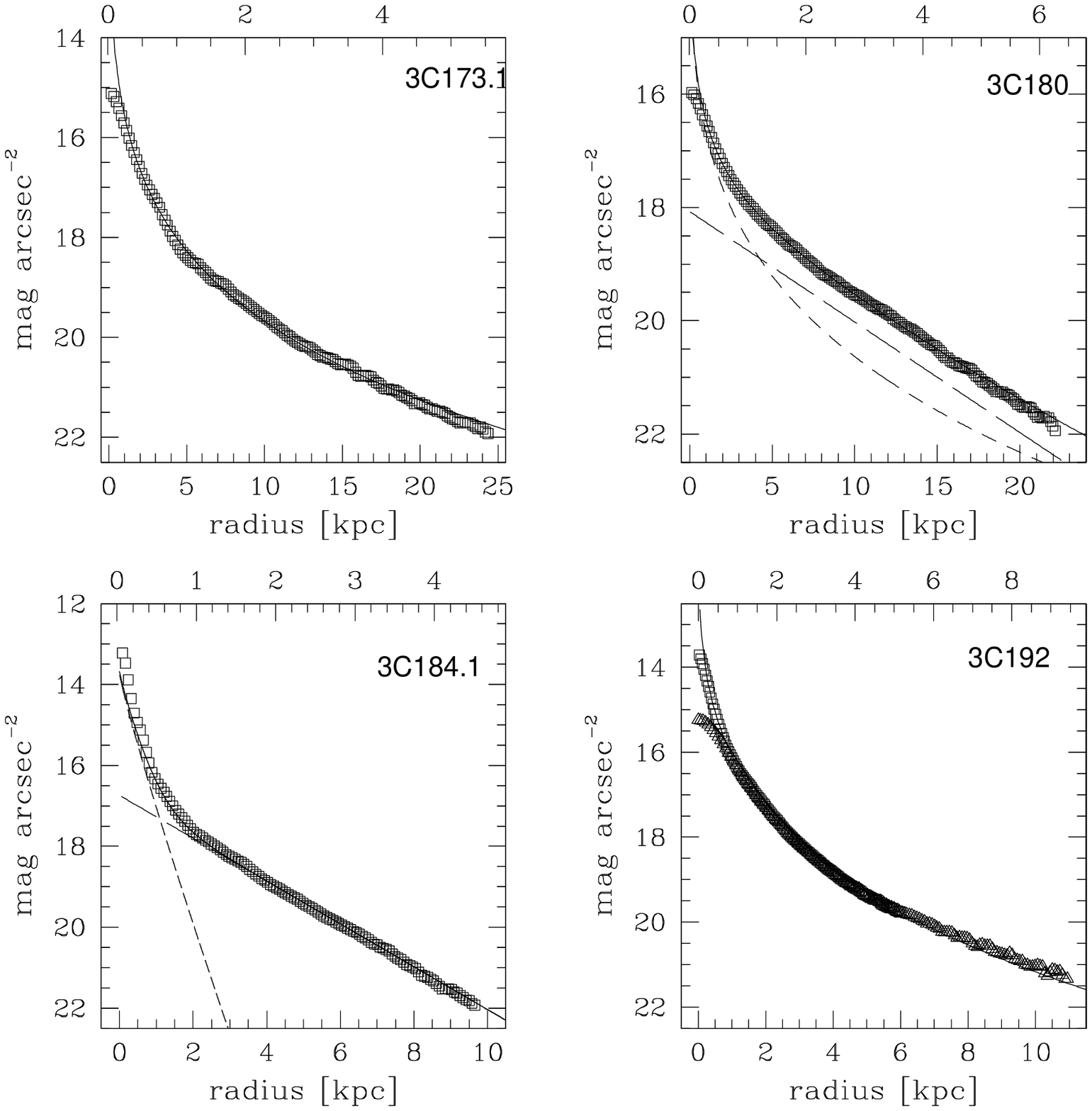}
\caption{Same as Fig. 1a.}
\end{figure}


\clearpage

\addtocounter{figure}{-1}
\addtocounter{subfig}{1}

\begin{figure}
\plotone{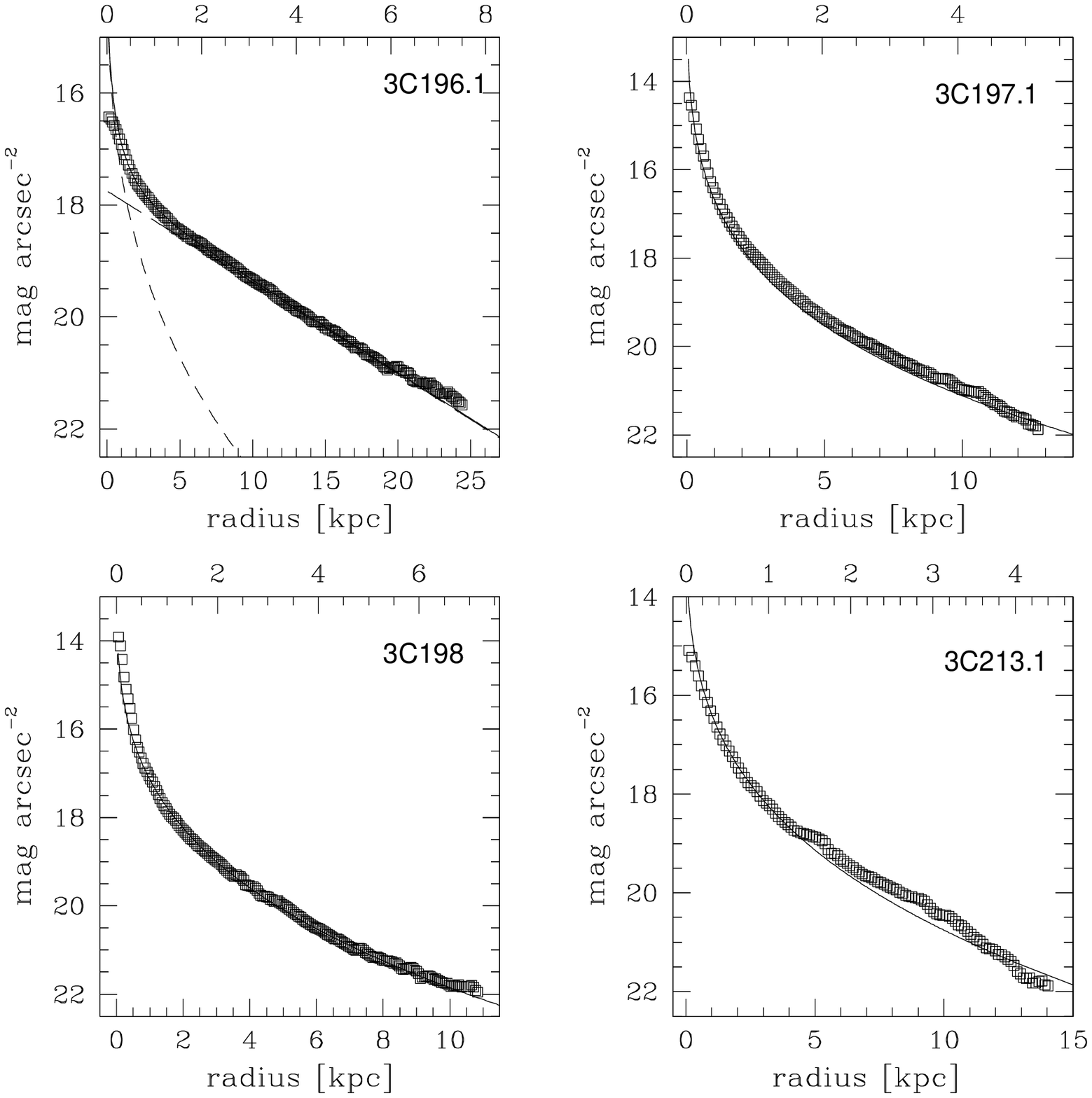}
\caption{Same as Fig. 1a.}
\end{figure}

\clearpage

\addtocounter{figure}{-1}
\addtocounter{subfig}{1}

\begin{figure}
\plotone{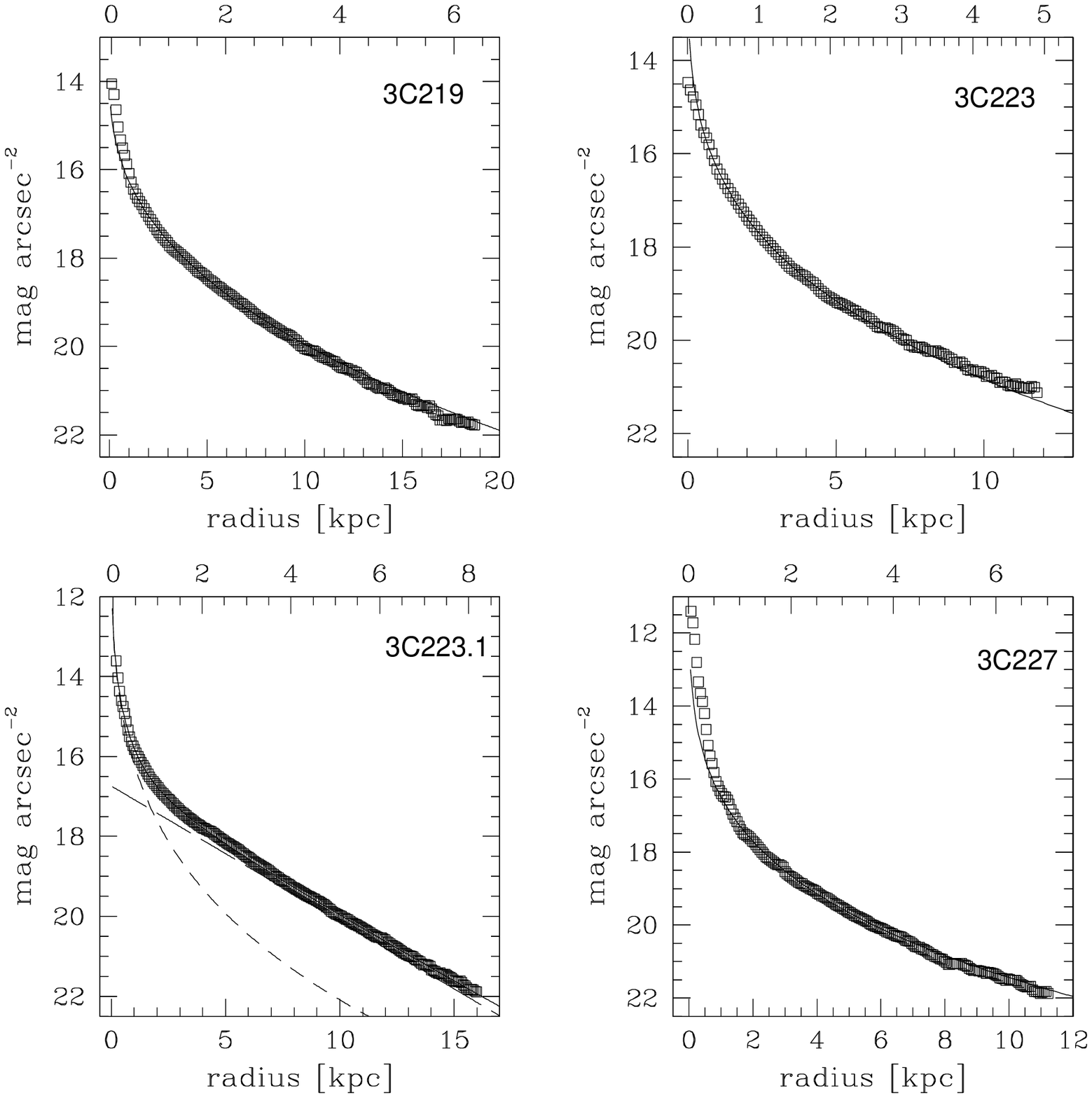}
\caption{Same as Fig. 1a.}
\end{figure}

\clearpage

\addtocounter{figure}{-1}
\addtocounter{subfig}{1}

\begin{figure}
\plotone{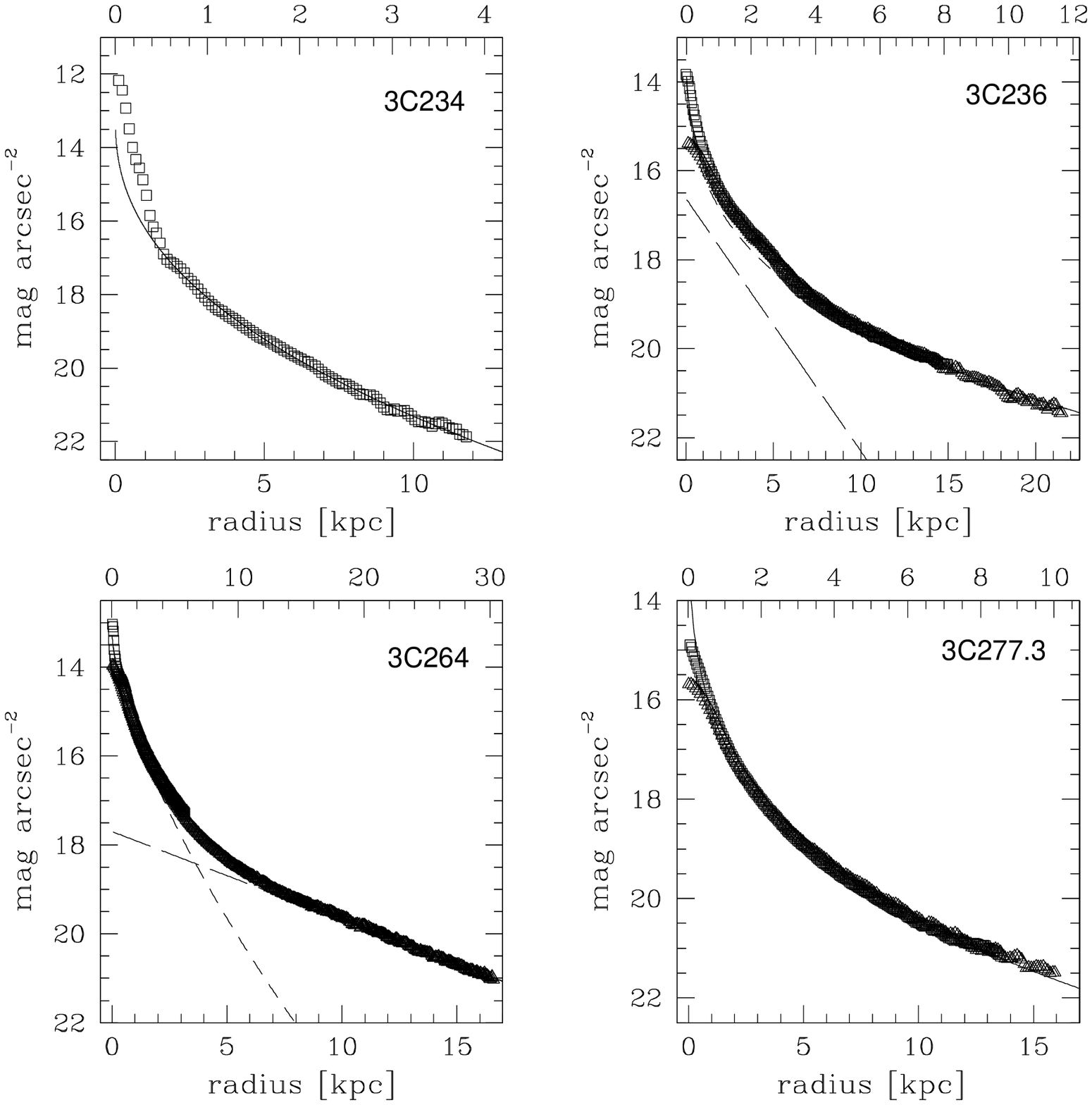}
\caption{Same as Fig. 1a.}
\end{figure}

\clearpage

\addtocounter{figure}{-1}
\addtocounter{subfig}{1}

\begin{figure}
\plotone{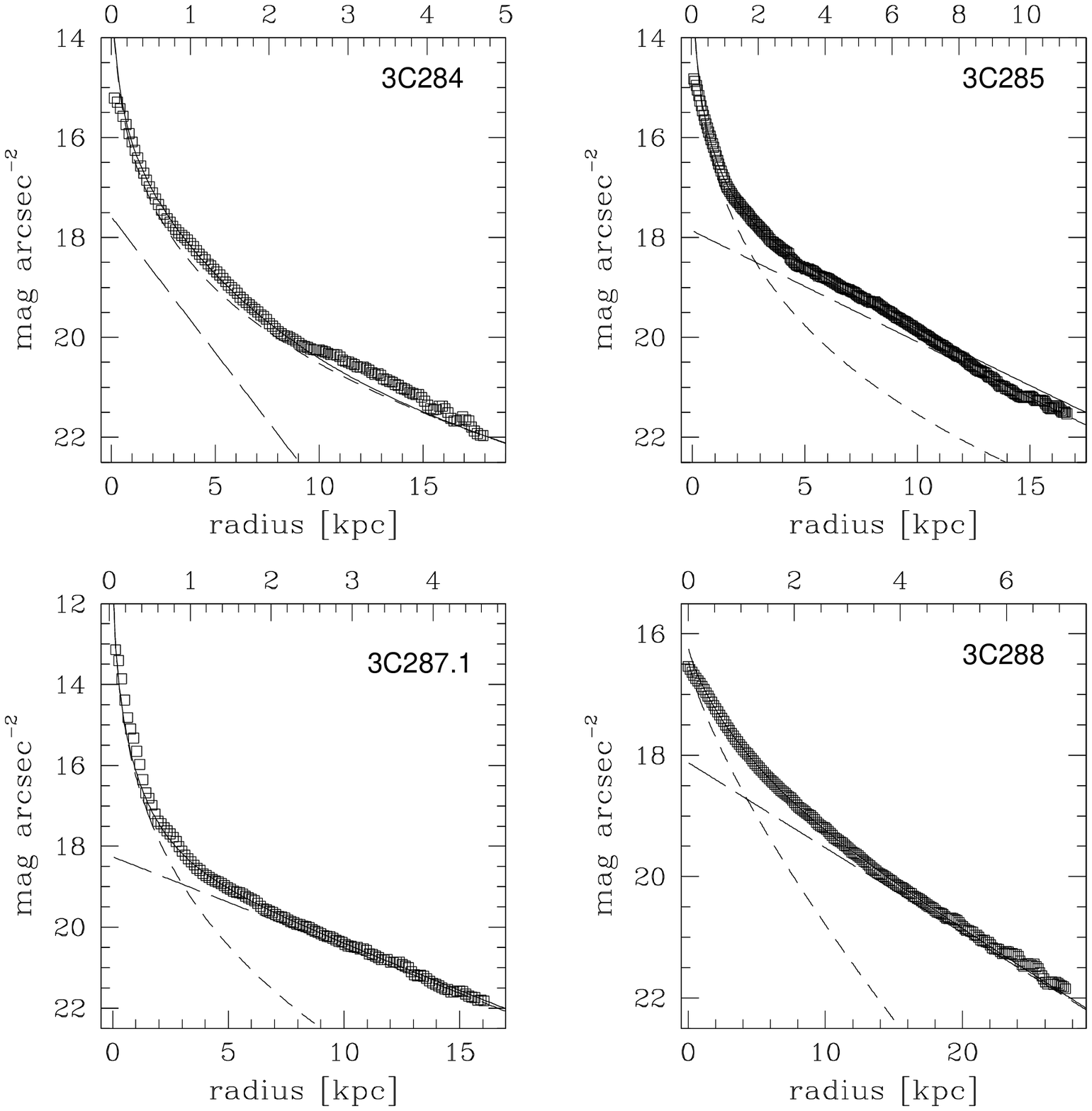}
\caption{Same as Fig. 1a.}
\end{figure}


\clearpage

\addtocounter{figure}{-1}
\addtocounter{subfig}{1}

\begin{figure}
\plotone{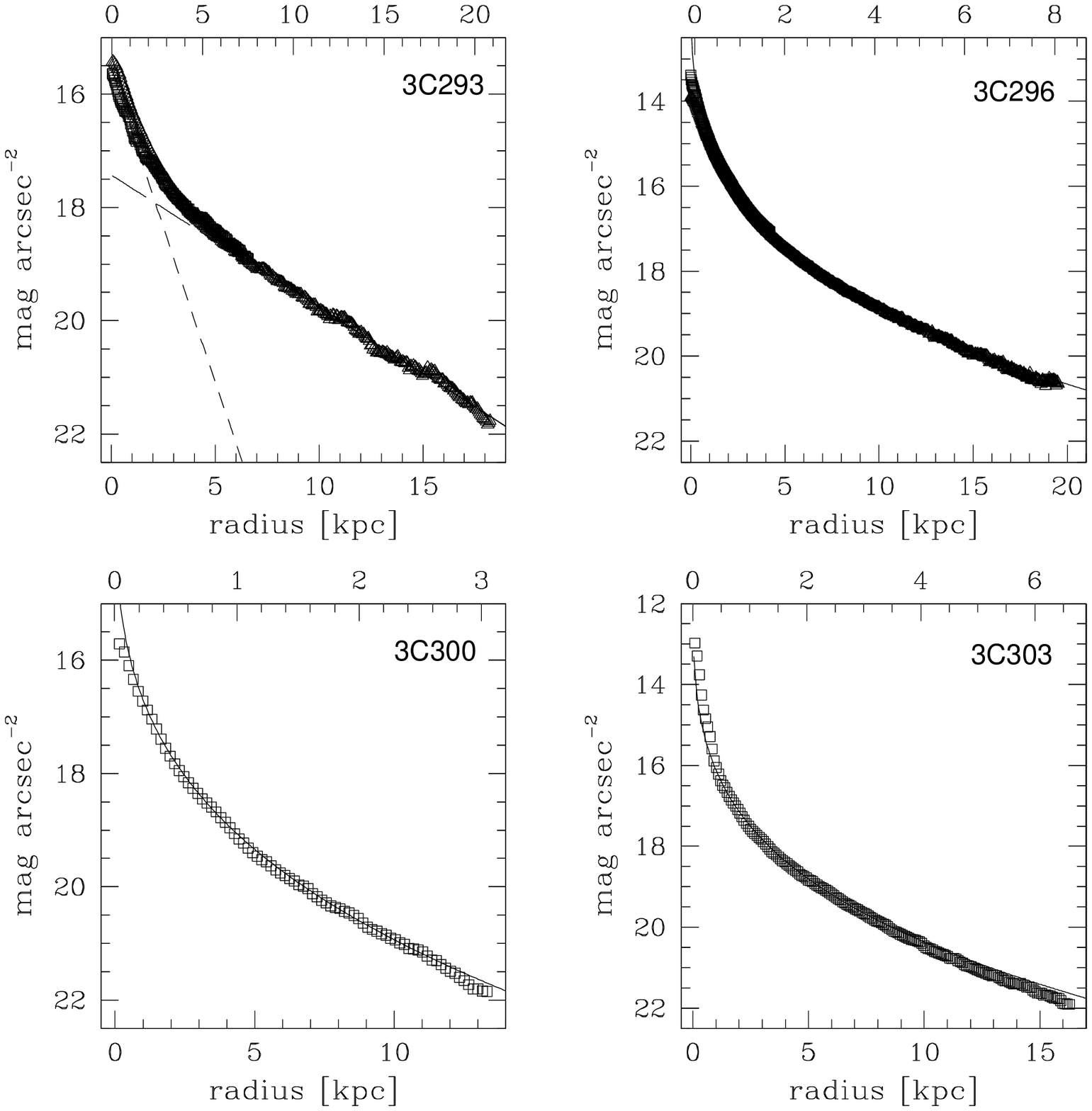}
\caption{Same as Fig. 1a.}
\end{figure}

\clearpage

\addtocounter{figure}{-1}
\addtocounter{subfig}{1}

\begin{figure}
\plotone{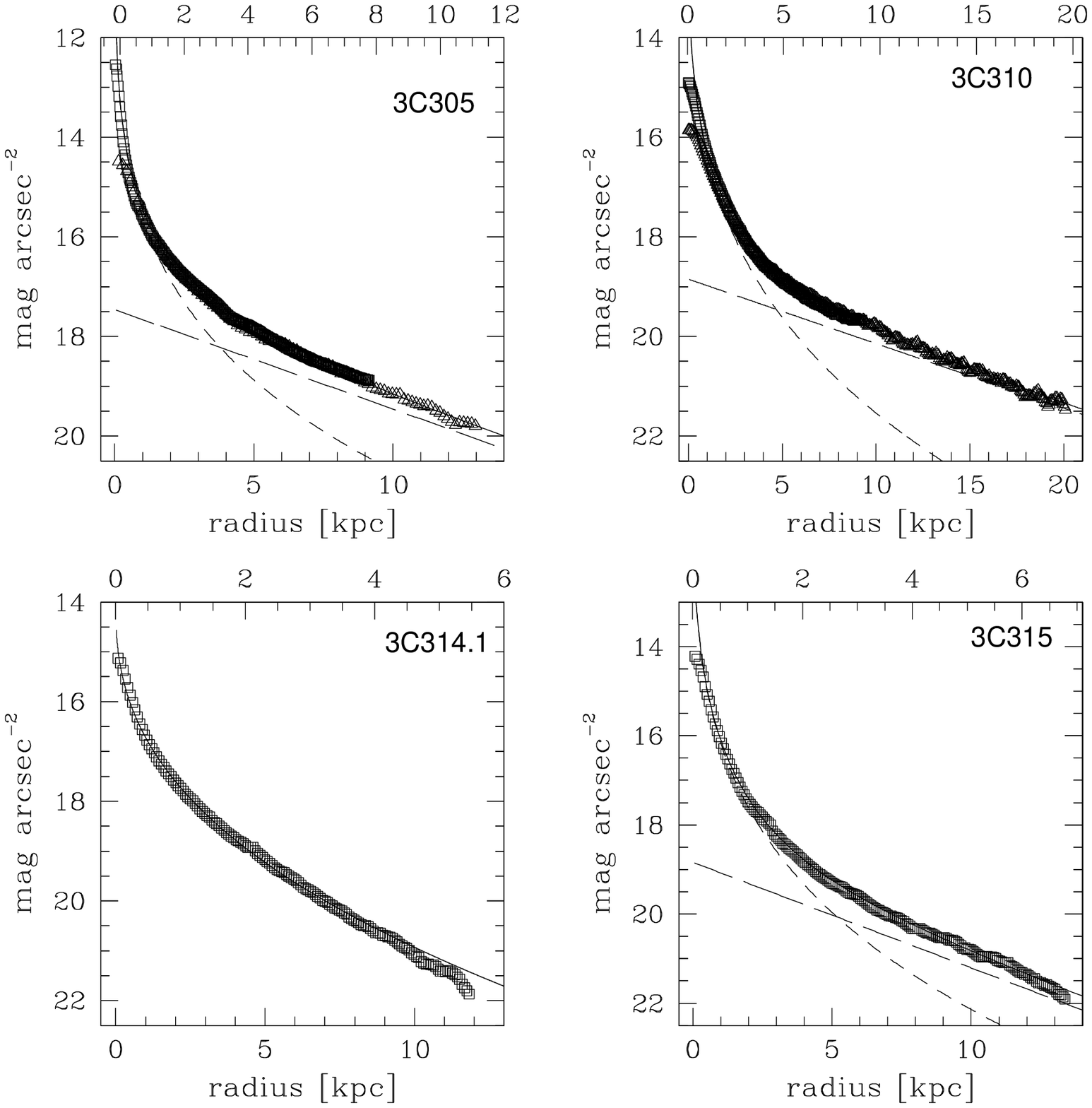}
\caption{Same as Fig. 1a.}
\end{figure}

\clearpage

\addtocounter{figure}{-1}
\addtocounter{subfig}{1}

\begin{figure}
\plotone{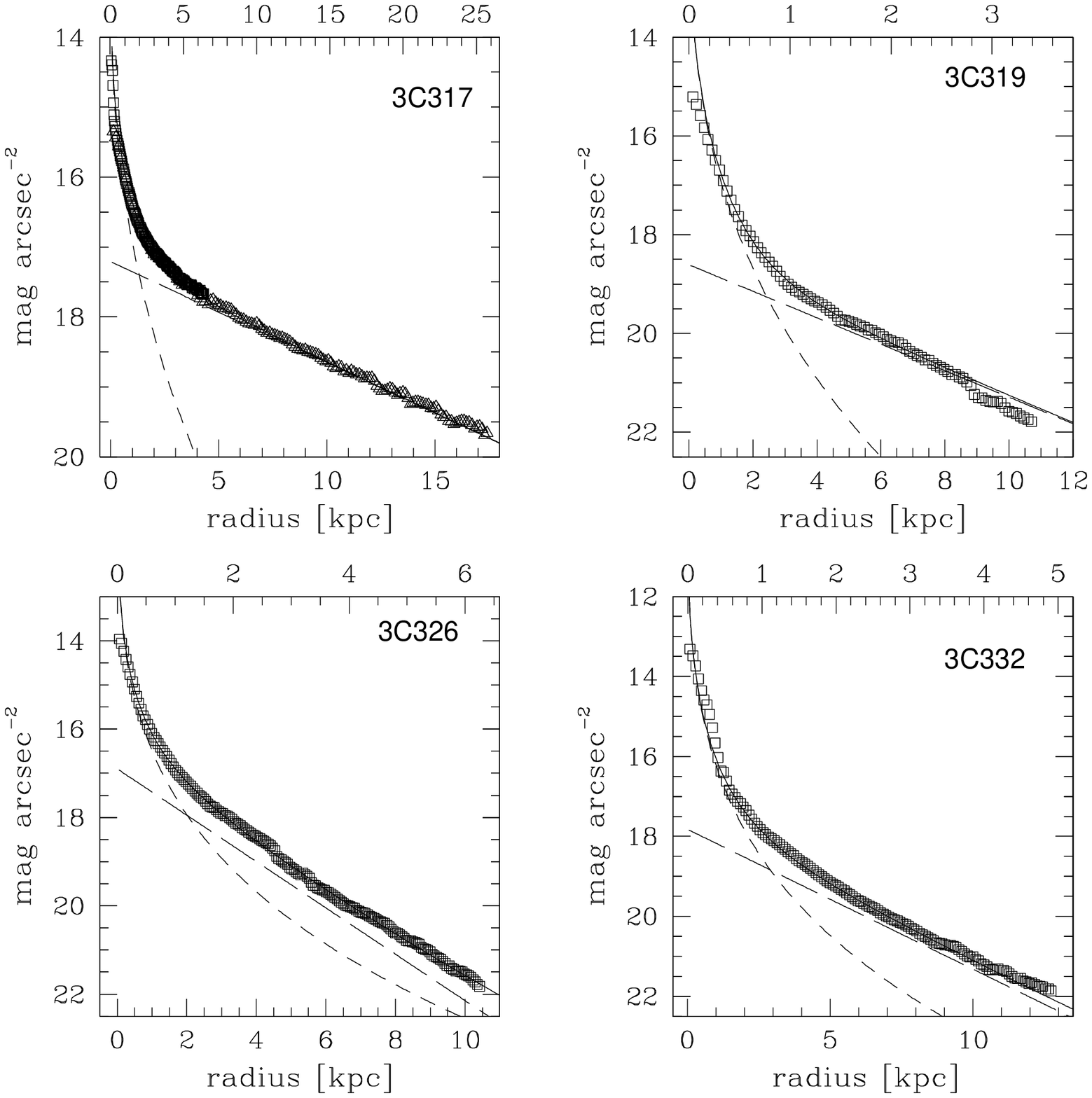}
\caption{Same as Fig. 1a.}
\end{figure}

\clearpage

\addtocounter{figure}{-1}
\addtocounter{subfig}{1}

\begin{figure}
\plotone{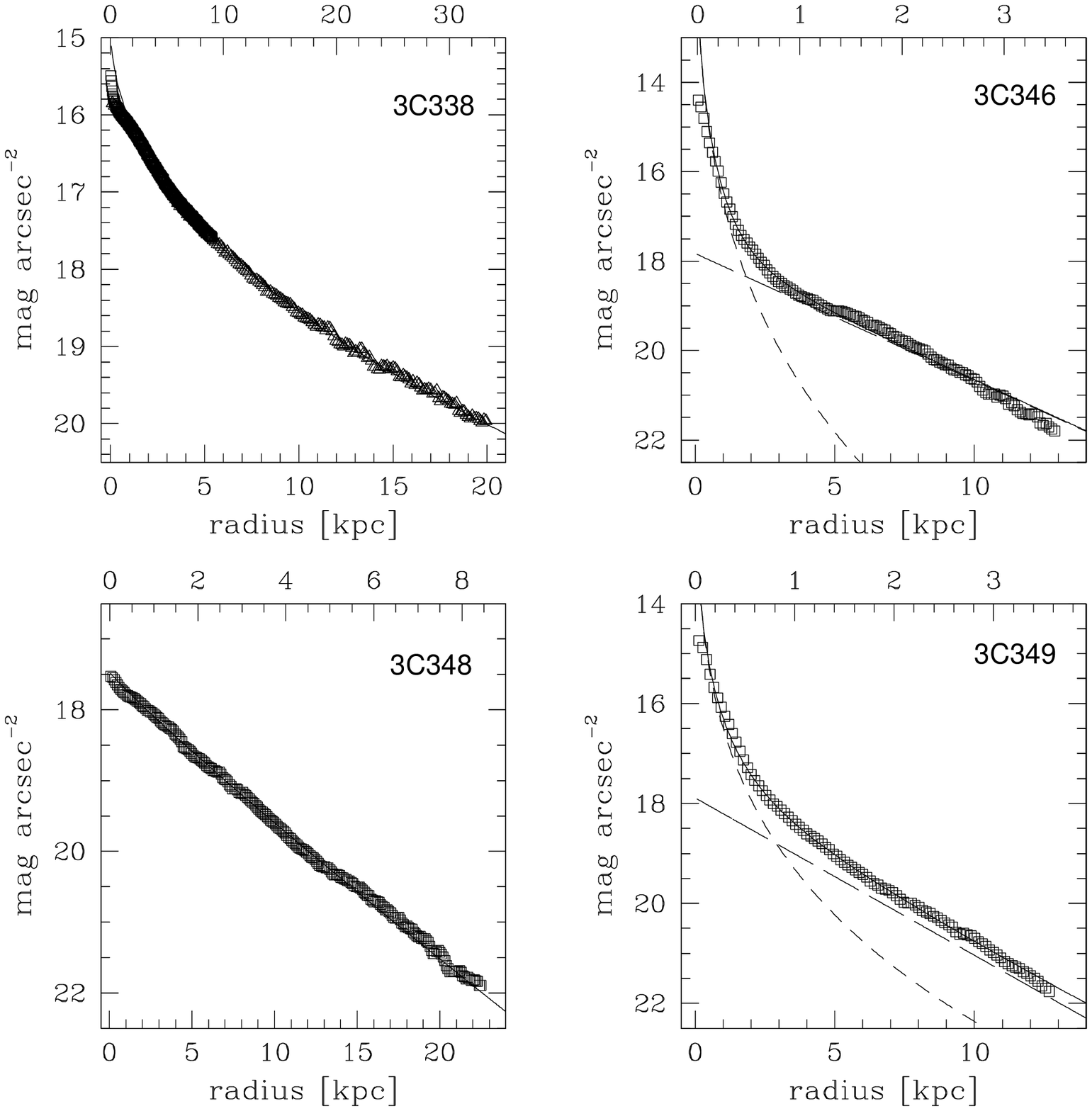}
\caption{Same as Fig. 1a.}
\end{figure}


\clearpage

\addtocounter{figure}{-1}
\addtocounter{subfig}{1}

\begin{figure}
\plotone{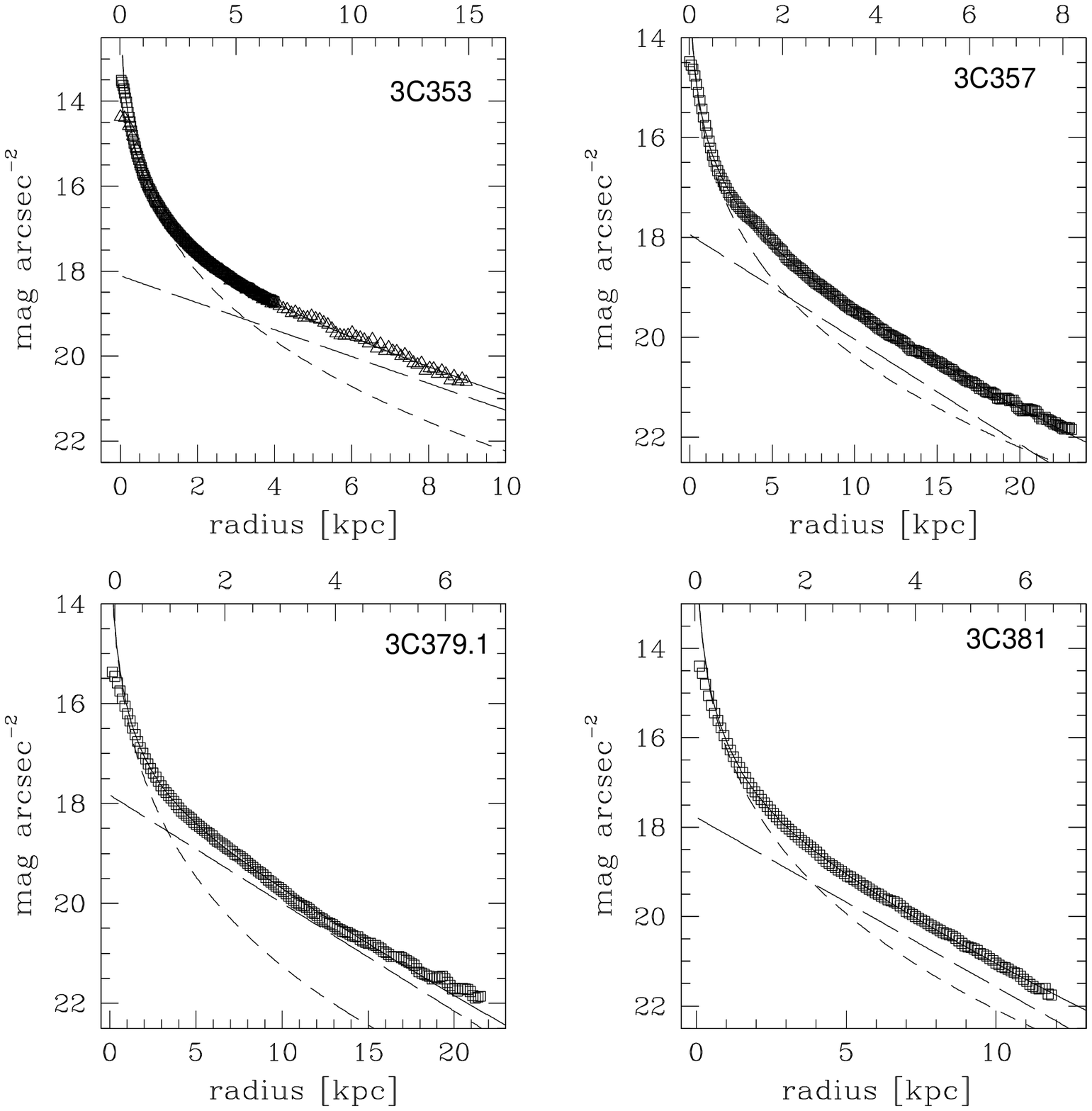}
\caption{Same as Fig. 1a.}
\end{figure}

\clearpage

\addtocounter{figure}{-1}
\addtocounter{subfig}{1}

\begin{figure}
\plotone{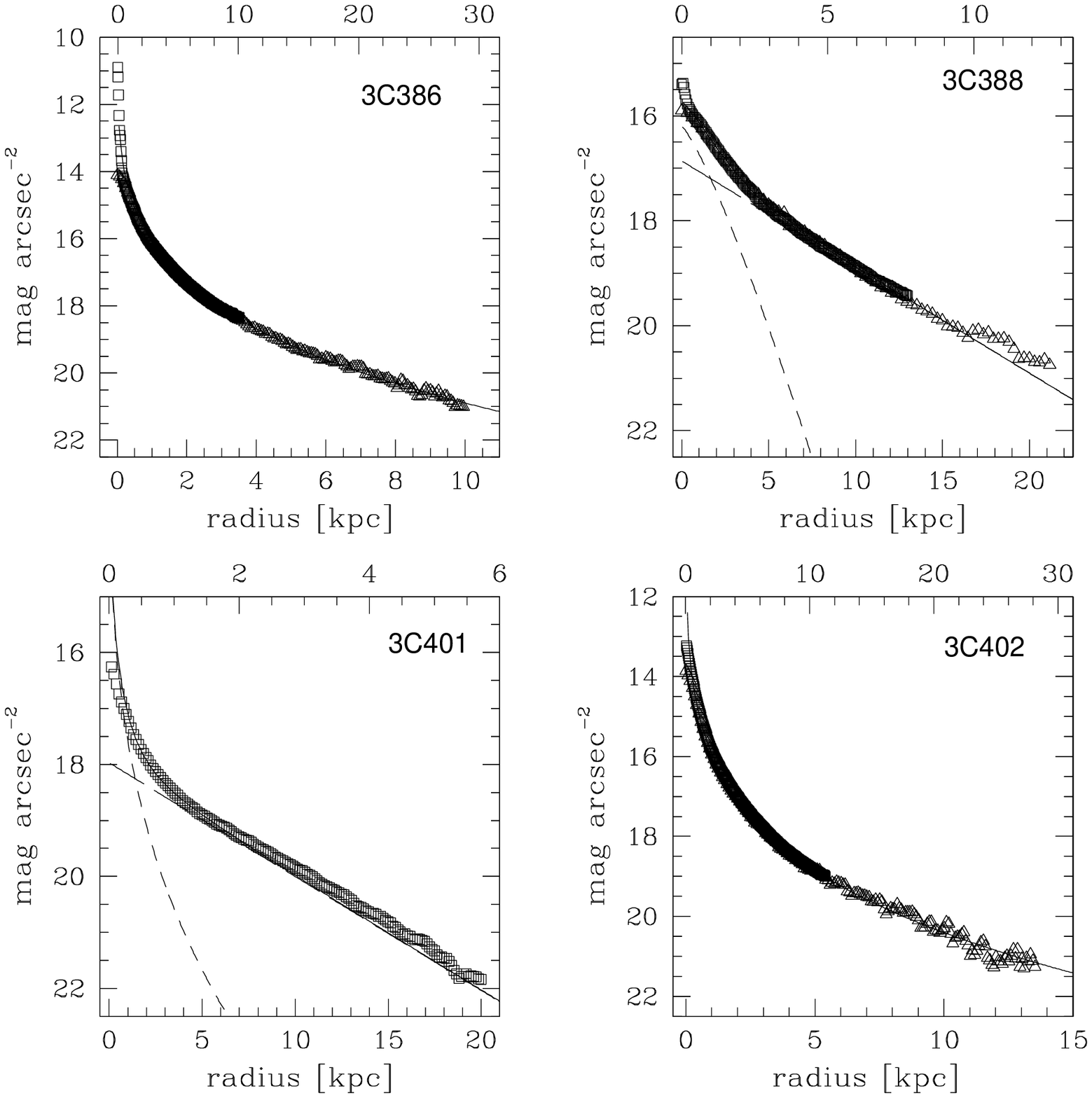}
\caption{Same as Fig. 1a.}
\end{figure}

\clearpage

\addtocounter{figure}{-1}
\addtocounter{subfig}{1}

\begin{figure}
\plotone{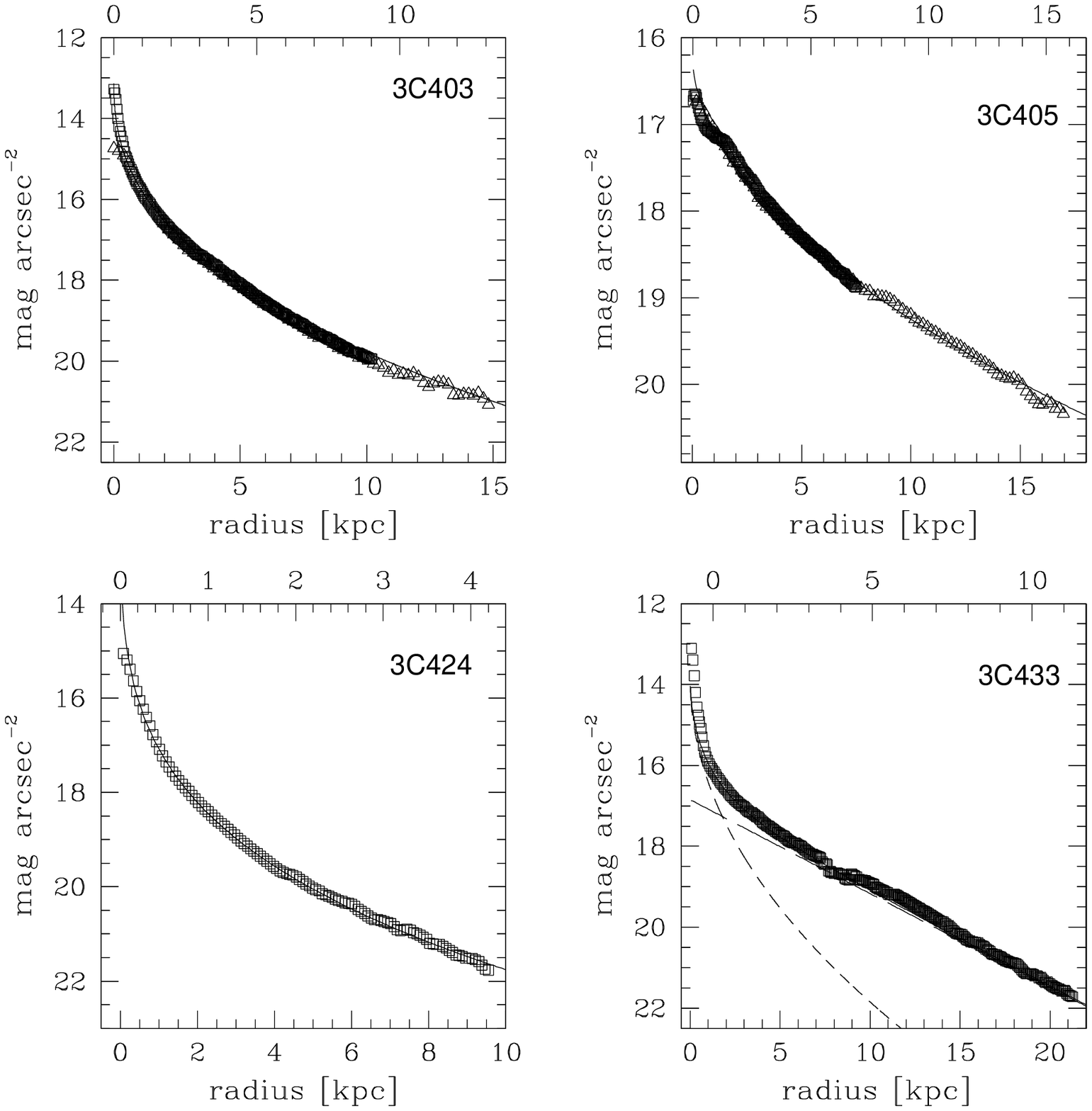}
\caption{Same as Fig. 1a.}
\end{figure}

\clearpage

\addtocounter{figure}{-1}
\addtocounter{subfig}{1}

\begin{figure}
\plotone{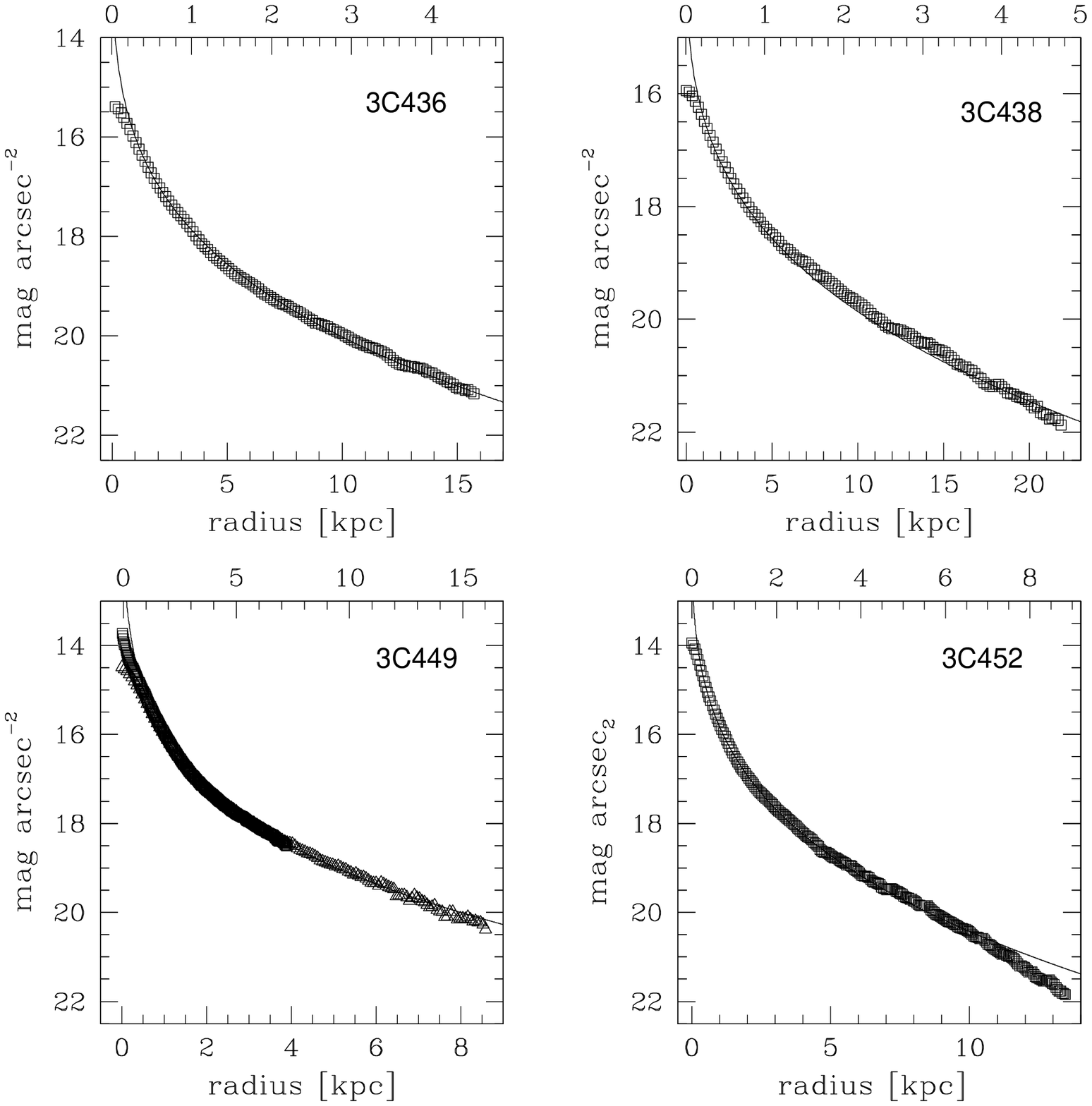}
\caption{Same as Fig. 1a.}
\end{figure}


\clearpage

\renewcommand{\thefigure}{\arabic{figure}}

\begin{figure}
\plotone{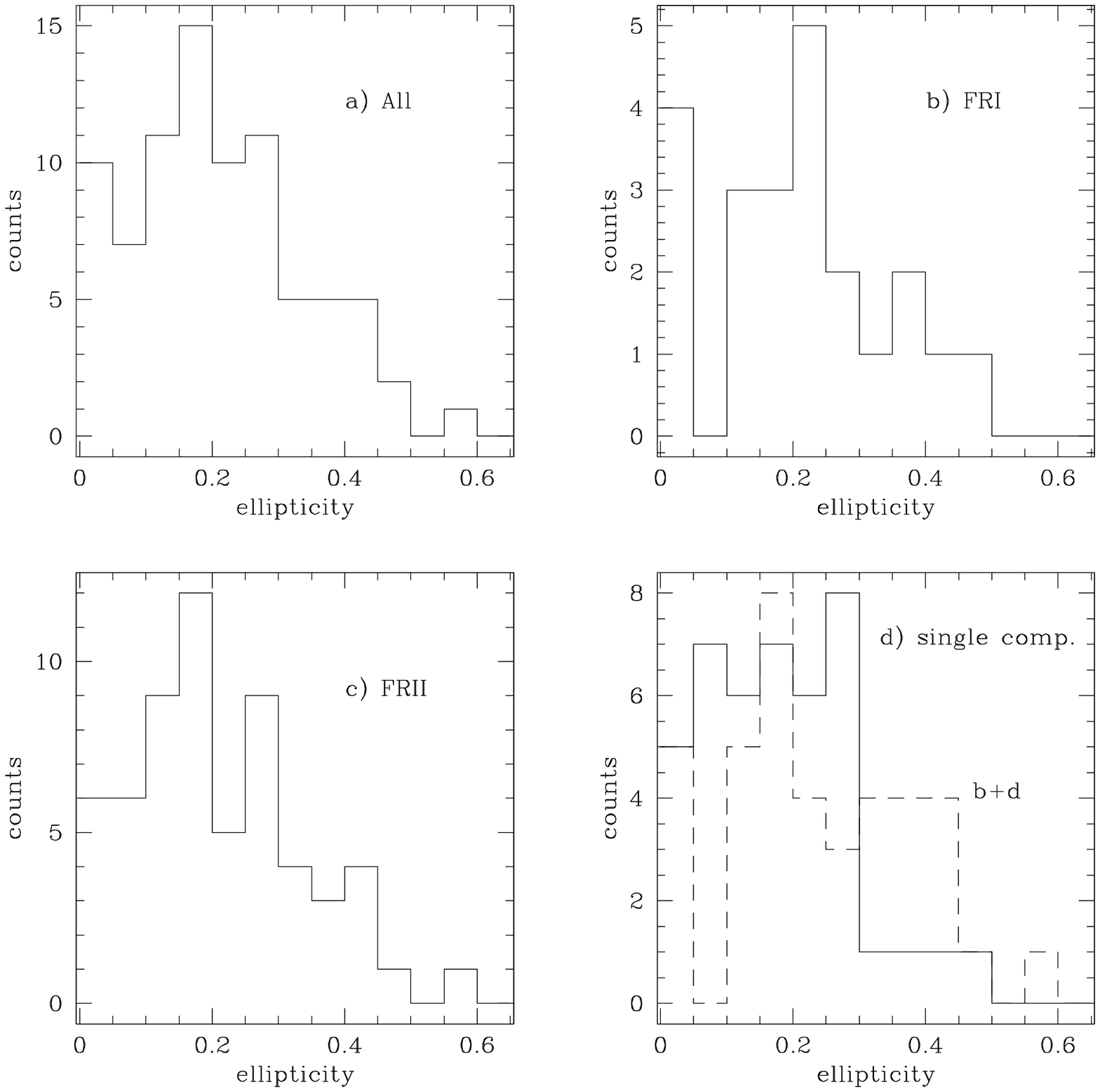}
\caption{Ellipticity distributions for: a) Whole sample, b) FR~I galaxies, c) FR~II galaxies, d) Galaxies
with single S\'ersic luminosity profile (solid line histogram) and for bulge+disk luminosity profile galaxies (dashed line histogram). }
\end{figure}

\clearpage
\begin{figure}
\plotone{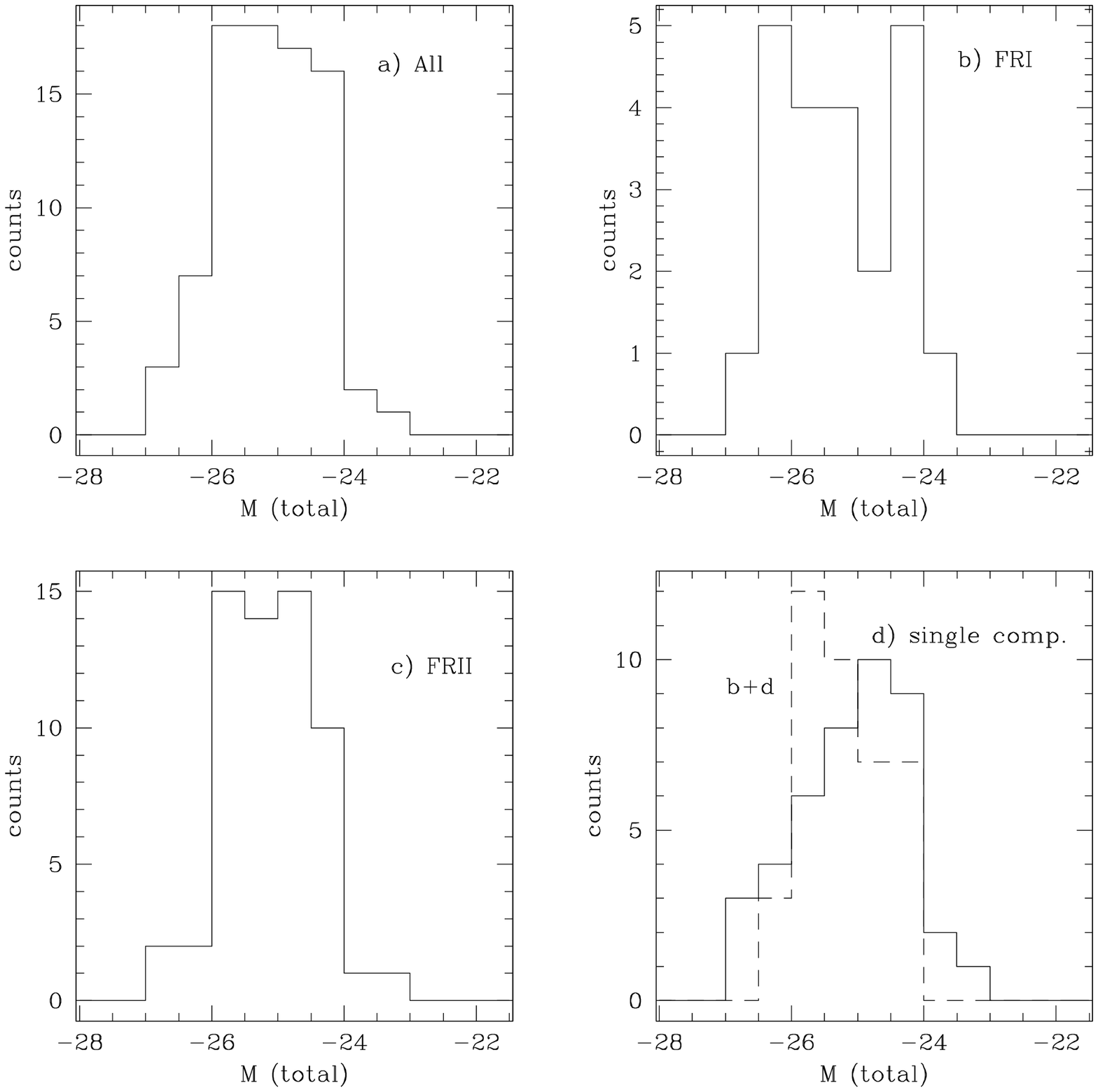}
\caption{Total luminosity distributions for: a) Whole sample, b) FR~I galaxies, c) FR~II galaxies, d) Galaxies
with single S\'ersic luminosity profile (solid line histogram) and for bulge+disk luminosity profile galaxies (dashed line histogram).}
\end{figure}

\clearpage
\begin{figure}
\plotone{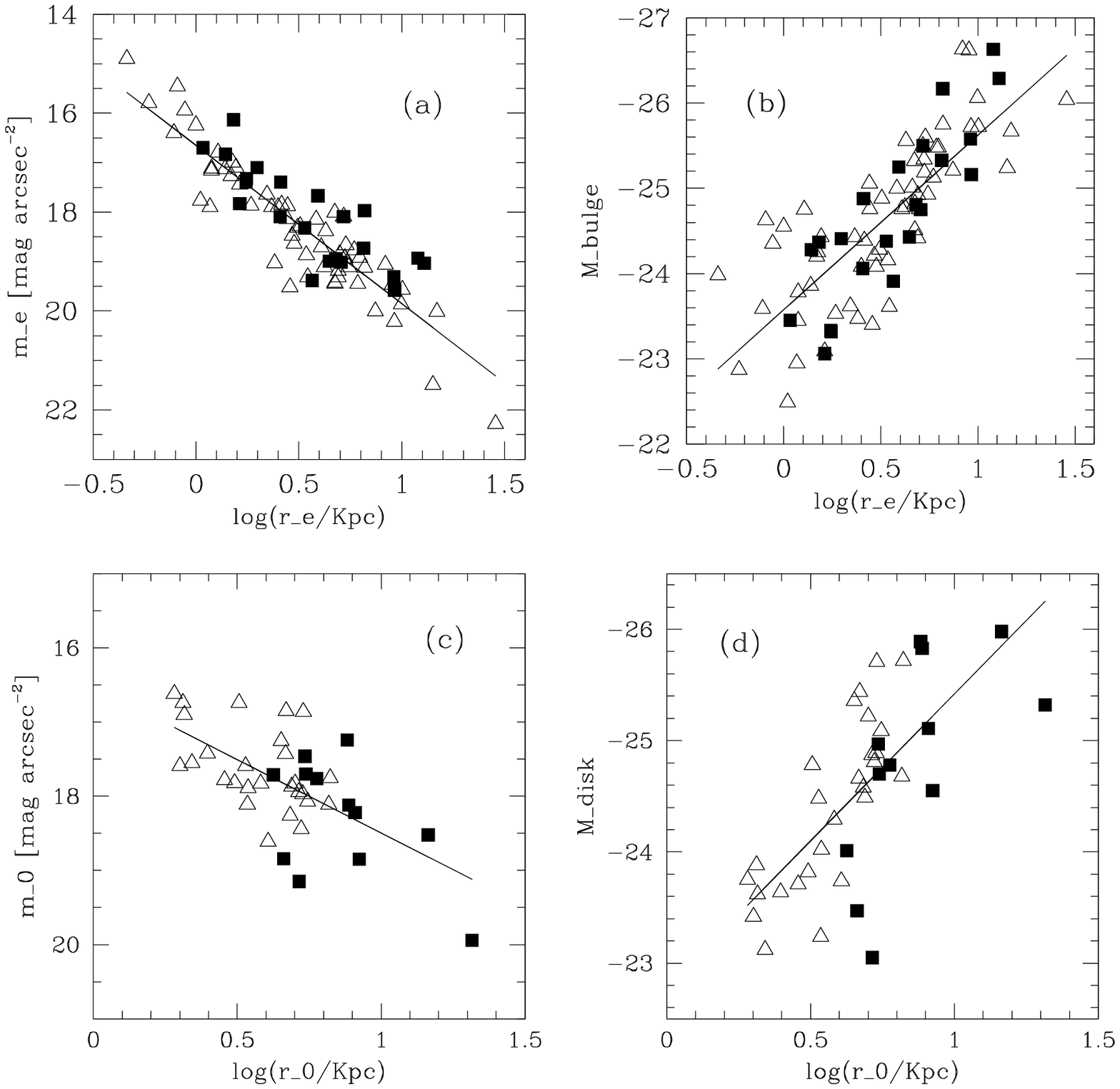}
\caption{Photometric correlations: a) Effective magnitude vs. log of the effective radius, b) bulge luminosity vs. log of the effective radius, 
c) central surface magnitude vs. log of the length scale, d) disk total luminosity vs. log of the length scale. Squares represent FR~I galaxies
while empty triangles represent FR~II galaxies. Lines show the result of the linear regression.}
\end{figure}

\clearpage
\clearpage

\begin{figure}
\plotone{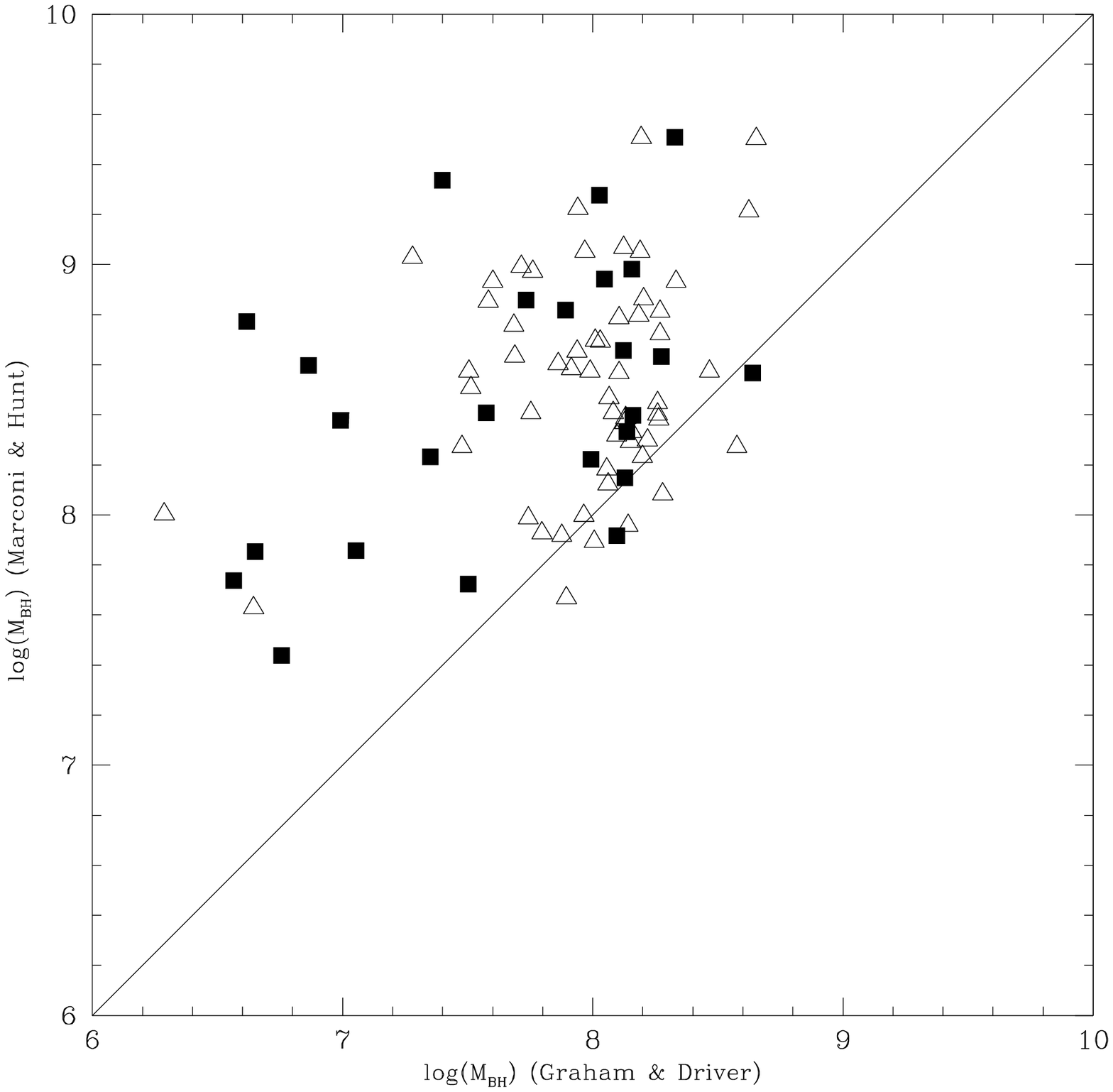}
\caption{Black hole masses calculated for the sample galaxies. X-axis show the results obtained using the Graham \& Driver (2006)
 relation while y-axis show masses obtained following the Marconi \& Hunt (2003) relation. Squares and empty triangles represent
  FR~I and FR~II galaxies respectively.
Solid line indicates the y = x function.}
\end{figure}

\clearpage
\begin{figure}
\plotone{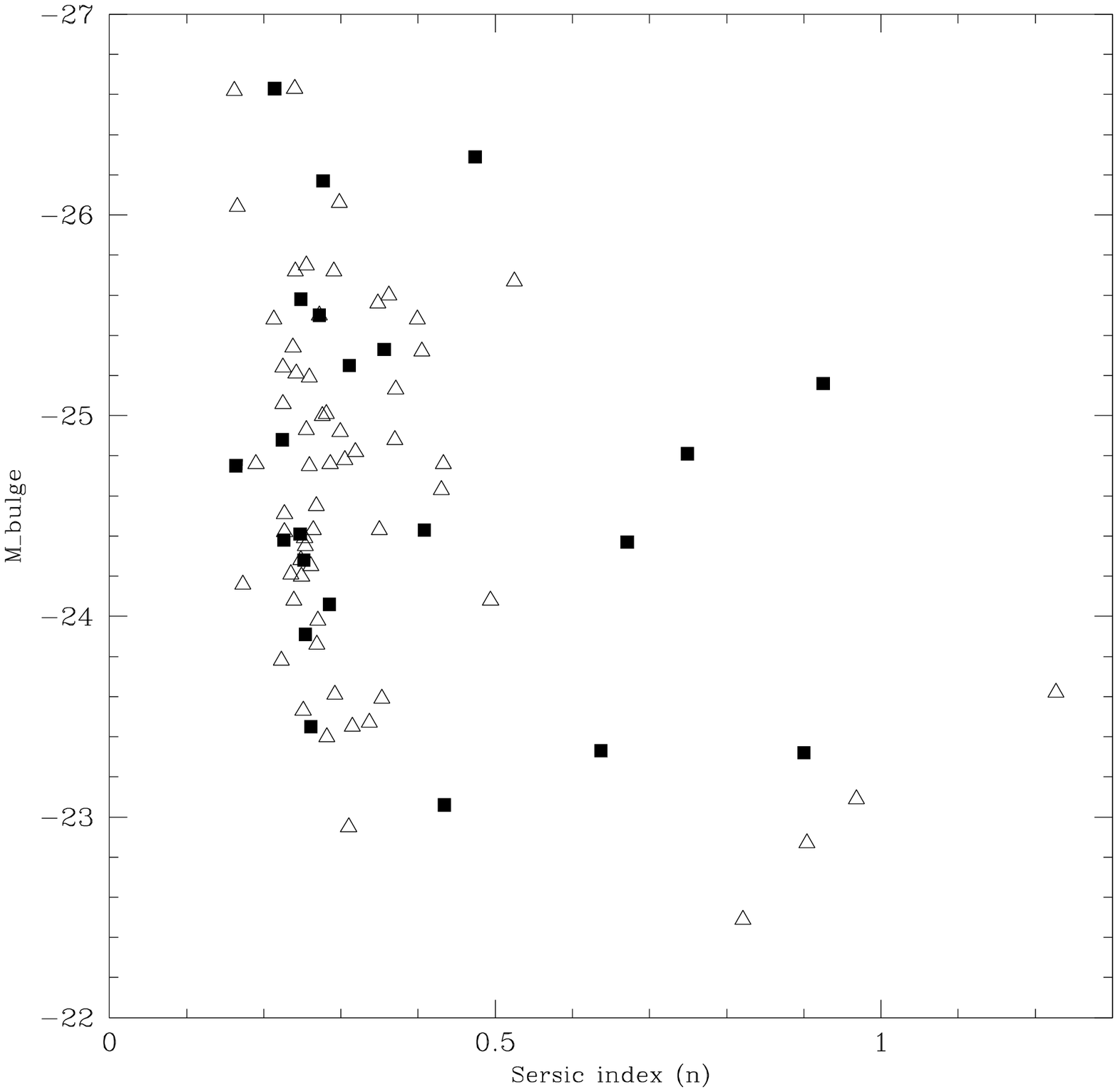}
\caption{Total bulge luminosity versus S\'ersic index $n$. Squares and empty triangles represent FR~I and FR~II galaxies respectively.}
\end{figure}


\end{document}